\documentclass[sigconf]{acmart}

\AtBeginDocument{%
  }

\copyrightyear{2025}
\acmYear{2025}
\acmConference[SenSys '25]{The 23rd ACM Conference on Embedded Networked Sensor Systems}{May 6--9, 2025}{Irvine, CA, USA}
\acmBooktitle{The 23rd ACM Conference on Embedded Networked Sensor Systems (SenSys '25), May 6--9, 2025, Irvine, CA, USA}

\newcommand{\greencheck}{{\color{green}\ding{51}}}
\newcommand{\redcross}{{\color{red}\ding{55}}}
\usepackage{balance} 
\usepackage{pifont}
\usepackage{circledsteps}
\usepackage{tabularray}
\usepackage{amsmath}
\usepackage{hhline}
\usepackage{booktabs}
\usepackage{tabularx}
\usepackage{graphicx}
\usepackage{siunitx}
\usepackage{float}
\usepackage{multirow}
\usepackage{tikz}
\usepackage{xcolor,colortbl}
\usepackage{titlesec}
\usepackage{url, hyperref}
\usetikzlibrary{tikzmark,calc}
\usepackage{xcolor}
\newcommand{\shepherd}[1]{\textcolor{black}{#1}}

\begin{document}

\title{Offload Rethinking by Cloud Assistance for Efficient Environmental Sound Recognition on LPWANs}


\author{Le Zhang\textsuperscript{*}, Quanling Zhao\textsuperscript{*}, Run Wang, Shirley Bian, Onat Gungor, Flavio Ponzina, Tajana Rosing}
\affiliation{
  \institution{University of California, San Diego}
  \city{La Jolla}
  \state{California}
  \country{USA}}
\email{{lez014, quzhao, ruw041, y1bian, ogungor, fponzina, tajana}@ucsd.edu}

\thanks{\textsuperscript{*}Both authors contributed equally to this research.}

\renewcommand{\shortauthors}{L. Zhang et al.}


\begin{abstract}

Learning-based environmental sound recognition has emerged as a crucial method for ultra-low-power environmental monitoring in biological research and city-scale sensing systems. These systems usually operate under limited resources and are often powered by harvested energy in remote areas. Recent efforts in on-device sound recognition suffer from low accuracy due to resource constraints, whereas cloud offloading strategies are hindered by high communication costs. In this work, we introduce \textit{ORCA}, a novel resource-efficient cloud-assisted environmental sound recognition system on batteryless devices operating over the Low-Power Wide-Area Networks (LPWANs), targeting wide-area audio sensing applications. We propose a cloud assistance strategy that remedies the low accuracy of on-device inference while minimizing the communication costs for cloud offloading. By leveraging a self-attention-based cloud sub-spectral feature selection method to facilitate efficient on-device inference, ORCA resolves three key challenges for resource-constrained cloud offloading over LPWANs: 1) high communication costs and low data rates, 2) dynamic wireless channel conditions, and 3) unreliable offloading. We implement ORCA on an energy-harvesting batteryless microcontroller and evaluate it in a real world urban sound testbed. Our results show that ORCA outperforms state-of-the-art methods by up to $80 \times$ in energy savings and $220 \times$ in latency reduction while maintaining comparable accuracy.

\end{abstract}




\maketitle

\section{Introduction}

\begin{figure}[!tp]
    \centering
    \includegraphics[width=\linewidth]{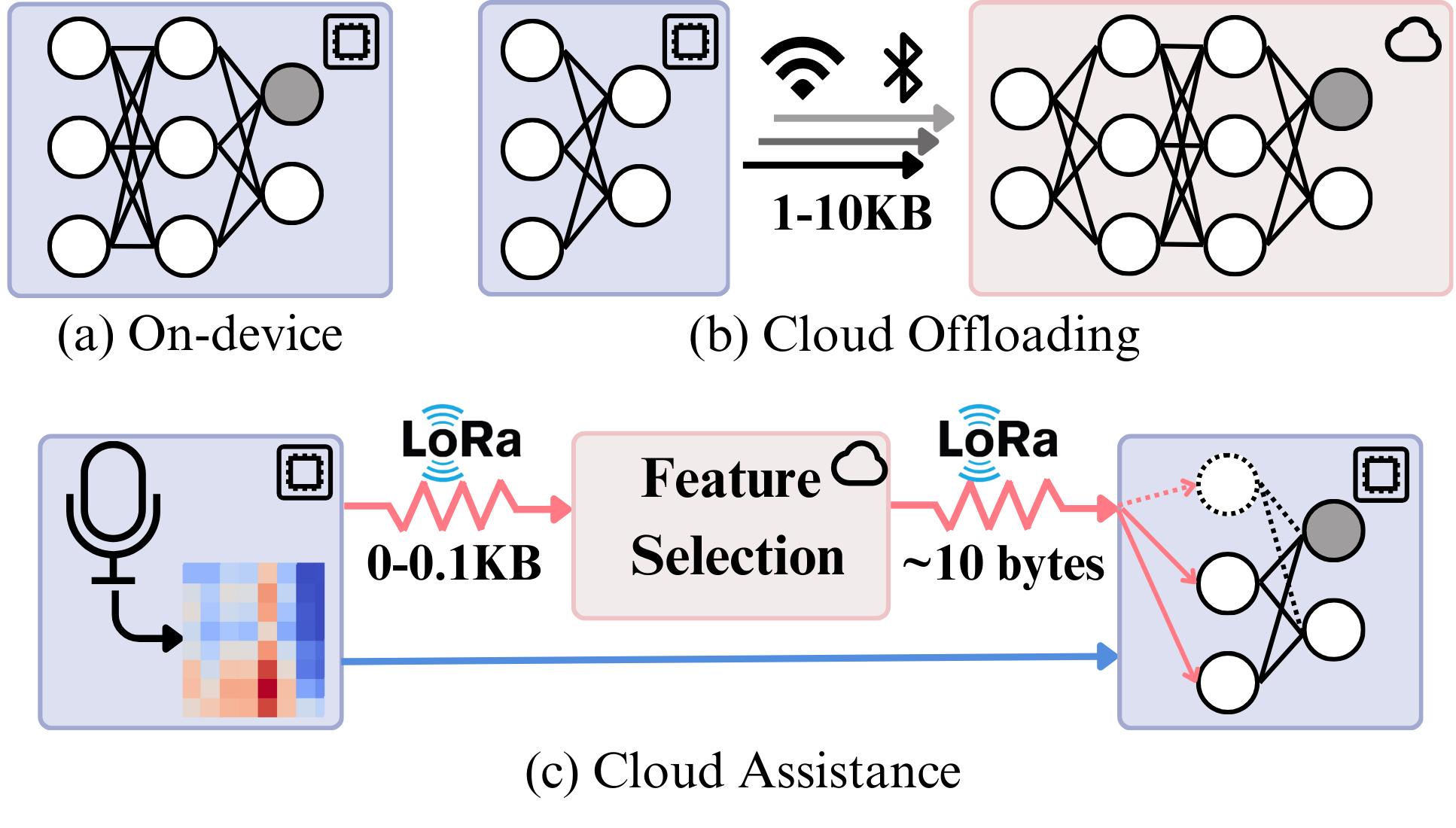}
    \vspace{-0.4cm}
    \caption{\shepherd{Comparisons of on-device inference, cloud offloading, and cloud assistance. }}
    \vspace{-0.4cm}
    \label{fig:three-methods}
\end{figure}

Recent advances in intelligent environmental monitoring highlight machine learning-based environmental sound recognition, leveraging ultra-low-power sensing in embedded systems which offers advantages over vision-based solutions~\cite{nirjon2013auditeur, lane2015deepear, ahn2024split}. These applications are usually deployed in remote and resource-constrained areas, such as forests, wildlife reserves, or urban environments, where energy-efficient operation is crucial for continuous, autonomous monitoring. They demonstrate potential in biological research, such as classifying bird vocalizations in the rainforest~\cite{salamon2016towards}, and in city-scale sensing applications, such as monitoring noise pollution and traffic flows~\cite{mohan2008nericell, rana2010ear, de2018paws}.

\shepherd{Despite their potential, edge devices' limited processing power and energy constraints make high-accuracy on-device predictions challenging~\cite{lee2019intermittent, gobieski2019intelligence, jeon2023harvnet, islam2020zygarde}. To overcome these limitations, recent efforts focus on edge-cloud collaboration~\cite{ahn2024split, huang2023rethink, yao2020deep, hojjat2024limitnet}. Compared to on-device learning, edge-cloud collaboration offloads inference tasks to the cloud server via wireless networks like Bluetooth LE, Wi-Fi, and LTE~\cite{ahn2024split, huang2023rethink, yao2020deep, hojjat2024limitnet}.  
This improves accuracy and reduces latency, but adds network dependency and energy overheads of costly wireless communication. Additionally, network coverage is a key factor in matching the diverse application scenarios. LTE provides wide coverage but is power-hungry for resource-constrained devices. In contrast, BLE and Wi-Fi are energy-efficient but have limited range, making them unsuitable for wide-area sensing applications. For instance, biologists deploy sensor nodes for acoustic bird species identification across vast rainforest areas~\cite{verma2024aviear}. Similarly, civil engineers and researchers use sensor nodes for large-scale acoustic sensing and classification in urban noise monitoring~\cite{tan2021extracting}. We find Low-Power Wide-Area Networks (LPWANs) ideal for these scenarios, balancing range and communication efficiency in edge-cloud collaborations. }

\shepherd{In this paper, we introduce a novel resource-aware cloud-assisted machine learning system, ORCA (\underline{O}ffload \underline{R}ethinking by \underline{C}loud \underline{A}ssistance), tailored for environmental sound recognition on ultra-low-power batteryless devices operating over LPWANs like LoRa networks. 
First, we leverage LPWANs to enable low-power, wide-area, and long-range edge-cloud collaboration for wide-area sound sensing applications. 
Second, we propose a novel \textit{cloud assistance} strategy for high-accuracy, resource-efficient, and LPWAN-adapted cloud offloading. In cloud assistance, instead of on-device inference as Figure~\ref{fig:three-methods}(a) or offloading by splitting the pipeline across wireless channels and generating inference results on the cloud as Figure~\ref{fig:three-methods}(b), ORCA keeps the entire inference pipeline on edge devices while leveraging the server for assistance. Specifically, the server selects the most important features uploaded by the edge devices. Here, we focus on identifying the most important frequency-domain features as previous studies show their effectiveness in environmental sound classification~\cite{phaye2019subspectralnet}. After that, the server provides the edge device with feature importance information through the downlink. This process is shown as the red arrows in Figure~\ref{fig:three-methods}(c). The edge devices can now use the given information to selectively process and perform inference on the most important features as the blue arrow in Figure~\ref{fig:three-methods}(c), thereby saving energy and time by scaling down the inputs and improving accuracy. As a result, the role of the server is now "assisting" edge devices for inference, avoiding edge devices being dependent on it. Specifically, ORCA tackles three key challenges of edge-cloud collaboration over LPWANs:}

\shepherd{
\textbf{(1) High communication costs and low data rate:} Though LPWANs like LoRa are “low-power”, wireless communication remains resource-intensive, e.g., 40$\times$ for uplink and 6$\times$ for downlink to the average power of on-device inference~\cite{mileiko2023run}. Additionally, LPWANs operate at a low data rate, e.g., 0.3-5.5 kbps in 125 kHz LoRa channel~\cite{bor2016lora}. As a result, previous data streaming approaches on BLE, Wi-Fi, or LTE~\cite{huang2023rethink, yao2020deep} are infeasible. To further reduce payload sizes, we propose transmitting low-resolution spectrograms for cloud processing. However, our preliminary studies indicate that while low-resolution spectrograms significantly reduce payloads compared to conventional cloud offloading methods, they also degrade cloud inference accuracy. On the other hand, compared to cloud inference, feature selection is generally more resilient to accuracy degradation with low-resolution spectrograms.
Therefore, we introduce a cloud-assisted approach that transmits low-resolution spectrograms to the cloud for feature selection rather than inference. ORCA implements this by feeding the low-resolution spectrogram into a pre-trained vision transformer and extracting feature importance masks from the self-attention maps.  }

\shepherd{\textbf{(2) Dynamic communication cost:} Wireless channel conditions are variable due to dynamic environments. To ensure reliable transmission, LPWANs adjust configurations like transmission power. Previous assumptions of constant wireless costs are impractical for energy-sensitive batteryless devices~\cite{yao2020deep, ahn2024split}. In ORCA, we dynamically optimize the "assistance resolution", the size of the spectrogram for uplink transmission and feature selection, to balance payload size and energy cost based on real-time communication feedback. For example, when communication cost is low, we select a larger size of spectrogram to improve feature extraction on the server and eventually increase the on-device classification accuracy. On the opposite, when communication cost is high, we switch to a smaller size of spectrogram for cloud assistance which result in lower classification accuracy.}

\shepherd{
\noindent \textbf{(3) Unreliable offloading:} Factors such as signal attenuation, interference, and environmental obstacles can lead to unreliable data transmission, resulting in packet loss and the need for retransmissions. This further challenges the reliability of the end-to-end offloading services. The previous server-dependent solution addresses this problem by resource-inefficient retransmission~\cite{hojjat2024limitnet}. In ORCA, when a packet loss occurs, we switch to local on-device inference only without cloud assistance, i.e., we keep the blue arrow only and ignore the red arrows in Figure~\ref{fig:three-methods}(c). This benefits from the design of maintaining the inference pipeline on-device, thereby saving the costs required for retransmission.}

\noindent
\shepherd{
\textbf{Application Scenarios:}
ORCA can operate over batteryless sensor networks in the wild, where distributed batteryless sensor nodes collect audio samples and perform resource-efficient cloud-assisted learning by connecting to an existing LPWAN gateway or a standalone edge server through wireless channels. As an aforementioned example~\cite{tan2021extracting}, batteryless sensor nodes powered by small solar panels can be strategically deployed in outdoor urban environments to monitor noise pollution. These nodes connect to established LoRa infrastructure and edge servers for resource-efficient cloud-assisted noise monitoring over hundred-meter to kilometer-scale distances. This setup enables continuous urban noise monitoring without battery replacements or maintenance, ensuring long-term, low-cost sustainability for smart city applications.}

\noindent
\textbf{Contributions: (1)} We propose a novel cloud assistance strategy for environmental sound recognition on resource-limited embedded systems and LPWANs. By leveraging vision transformer architecture for subspectral feature selection and efficient on-device spectral encoding convolutional neural networks (CNNs), ORCA achieves 2.5–12.5 p.p. accuracy improvements on five public environmental sound datasets.

\noindent
\textbf{(2)} We provide optimized assistance resolution strategies under dynamic communication costs and packet losses to ensure resource-efficient cloud assistance in dynamic environments.

\noindent
\textbf{(3)} We build a real-world testbed to evaluate accuracy, payload size, energy consumption, inference latency, and system overhead using collected LoRa traces. ORCA achieves up to 80$\times$ energy savings and 220$\times$ latency reduction, while maintaining accuracy comparable to the state-of-the-art methods. We release an open-source LoRa library for batteryless computing on the ultra-low-power MSP430 microcontroller\footnote{Library available at: \url{https://github.com/lezhangleonard/ORCA}}.

\definecolor{Gray}{gray}{0.95}
\definecolor{LightCyan}{rgb}{0.9,1,1}

\setlength{\arrayrulewidth}{0.4mm}
\begin{table*}[ht]
\centering
{\fontsize{9pt}{11pt}\selectfont
\setlength{\tabcolsep}{2pt}
\resizebox{\textwidth}{!}{%
\begin{tabular}{lllllccccc}
\hline
Prior Works & Edge Optimize & Wireless & Payload Data & Compression & Payload Size & Comm-adapt & ULP & Reliable & Testbed\\
\hline
\rowcolor{Gray}
DeepCOD~\cite{yao2020deep}           & Cloud offloading & Wi-Fi,LTE & Latent feature & Autoencoder & 0.1-5KB & \greencheck & \redcross & \redcross & \greencheck\\
FLEET~\cite{huang2023rethink}        & Early exits & BLE & Latent feature & Autoencoder & 2-9KB & \greencheck & \redcross & \redcross  & \greencheck \\
\rowcolor{Gray}
SEDAC~\cite{ahn2024split}            & Cloud offloading & BLE & Melspectrogram & Clip selection &  N/A & \redcross & \greencheck & \redcross & \greencheck \\
CACTUS~\cite{rastikerdar2024cactus}  & Micro-classifier & N/A & Model weights  & Class similarity & 11.5-43.9MB & \redcross & \redcross & \redcross & \greencheck \\
\rowcolor{Gray}
LimitNet~\cite{hojjat2024limitnet}   & Cloud offloading & LPWANs  & Latent feature & Gradual score & 0.3-3.2KB & \greencheck & \redcross & \redcross & \redcross\\
\rowcolor{LightCyan}
\textbf{ORCA}                                 & \textbf{Cloud assistance} & \textbf{LoRa} & \textbf{Spectrogram} & \textbf{Low resolution} & \textbf{0-0.1KB} & \greencheck & \greencheck &  \greencheck &  \greencheck \\
\hline
\end{tabular}
}
}
\caption{\shepherd{Comparisons of prior works on cloud offloading.}}
\label{tab:sota}
\vspace*{-0.8cm}
\end{table*}

\section{Background and Related Works}
\label{sec:background-and-related-works}

\subsection{Sound Recognition}
\label{sec:sound-recognition}
\noindent
\textbf{Audio Processing:} Audio signals are high-dimensional time-series data with complex patterns and temporal dependencies, making analysis challenging. 
\shepherd{Recent work in efficient sound classification uses Short-Time Fourier Transform (STFT) with CNNs or self-attention models to extract key time-domain features~\cite{monjur2023soundsieve, ahn2024split}. These methods rely only on time-domain features and often perform poorly in environmental sound recognition. Additionally, STFT introduces overhead and lacks flexibility for selective frequency analysis, as it provides uniform resolution across bands. On the another hand, recent studies show that sub-spectral features in the frequency domain are effective for acoustic scene classification, as environmental sounds often have unique frequency distributions~\cite{phaye2019subspectralnet, chang2021subspectral}. Therefore, we explore using sub-spectral features for resource-efficient sound recognition.}

\noindent
\textbf{Wavelet Transform:}
The wavelet transform provides a flexible approach for sub-spectral feature extraction, enabling analysis of non-stationary audio signals by decomposing them with orthonormal basis functions in $L^2$ space~\cite{kronland1987analysis, daubechies1990wavelet, lambrou1998classification}. Compared to STFT, the wavelet transform provides better localization in both time and frequency domains~\cite{ahn2024split, monjur2023soundsieve}. The discrete wavelet transform (DWT)~\cite{shensa1992discrete} uses a high-pass filter and a low-pass filter at various levels to break the signal into detailed and approximate coefficients.
DWT focuses on low-frequency decomposition, while the wavelet packet transform (WPT)~\cite{laine1993texture} provides a more detailed and flexible decomposition by analyzing both low and high frequencies, offering a comprehensive signal representation across all frequency bands. Recent research has explored integrating wavelet transforms with deep learning for sound recognition~\cite{wu2020time, qian2017wavelets, nakamura2020time, frusque2022learnable}. \shepherd{However, since frequency-domain features are not uniformly distributed, fine-tuning resolution is essential for efficient classification.}

\subsection{Cloud Offloading}
\noindent
\textbf{Offloading Solution:}
\shepherd{
Cloud offloading allocates partial computations to a cloud server, reducing the workload on edge devices. We summarize recent works in Table~\ref{tab:sota}. DeepCOD~\cite{yao2020deep}, SEDAC~\cite{ahn2024split}, and LimitNet~\cite{hojjat2024limitnet} split the inference pipeline across wireless channels. FLEET~\cite{huang2023rethink} adds early exits for resource savings. CACTUS~\cite{rastikerdar2024cactus} exchanges models for context-aware inference. We refer to these solutions as “cloud-dependent,” as their functionalities rely on the server. However, collaboration over LPWANs is limited by resource constraints and unreliable channels, making cloud-dependent strategies impractical. First, cloud-dependent approaches ignore edge resource constraints, offloading even basic inference task to the server. In contrast, empirical measurements show LoRa radios consume up to 40$\times$ more power for uplink and 6$\times$ for downlink than on-device computation~\cite{mileiko2023run}. Additionally, unreliable channels make consistent offloading infeasible. Data packets may be lost or corrupted during transmission, and edge devices may lack resources for retransmission. To tackle these challenges, we propose a novel \textit{cloud assistance} solution, contrasting with cloud-dependent strategies. In ORCA, the complete inference pipeline remains on edge devices for inference reliability. The server assists edge devices in identifying key input features, reducing their computational load.}


\noindent
\textbf{Data Compression:}
\shepherd{Data compression mitigates high communication costs under low bitrates. DeepCOD~\cite{yao2020deep} and FLEET~\cite{huang2023rethink} use autoencoders to compress latent embeddings dynamically. SEDAC~\cite{ahn2024split} and LimitNet~\cite{hojjat2024limitnet} select key input features for communication efficiency. LPAI~\cite{jing2022lpai} uploads selected audio clips for cloud training and requires downloading updated models. These approaches focus on transmitting compressed embeddings, input features, or model weights essential for cloud inference or training, aligning with the cloud-dependent strategy. However, large payloads are unsuitable for low-bitrate networks and resource-constrained devices. We propose an alternative approach to transmit smaller data and leverage complex server-side processing for cloud assistance.} In ORCA, we use low-level feature abstractions like low-resolution audio spectrograms as a compressed form of data, significantly reducing transmission size. Despite the small size, the low-resolution feature abstraction aids the inference process substantially, as we will demonstrate later in this paper. We draw an analogy to the bounding box mechanism used in regional proposal networks (RPN)~\cite{ren2015faster} and YOLO~\cite{redmon2016you}, which utilize low-resolution images to propose potential regions of interest and perform efficient classification within these bounded regions. However, SEDAC~\cite{ahn2024split} indicates that it is infeasible to deploy RPN for audio tasks on resource-constrained devices due to the high computational demands and fundamental differences between the vision and audio domains. Thus, we conclude that there is a clear need for an audio-adaptive, cloud-assisted, and communication-efficient collaborative learning framework to enable resource-efficient environmental sound recognition.


\section{Preliminary Study}
\label{sec:preliminary-study}

\begin{figure}[tp]
    \centering
    \includegraphics[width=\linewidth]{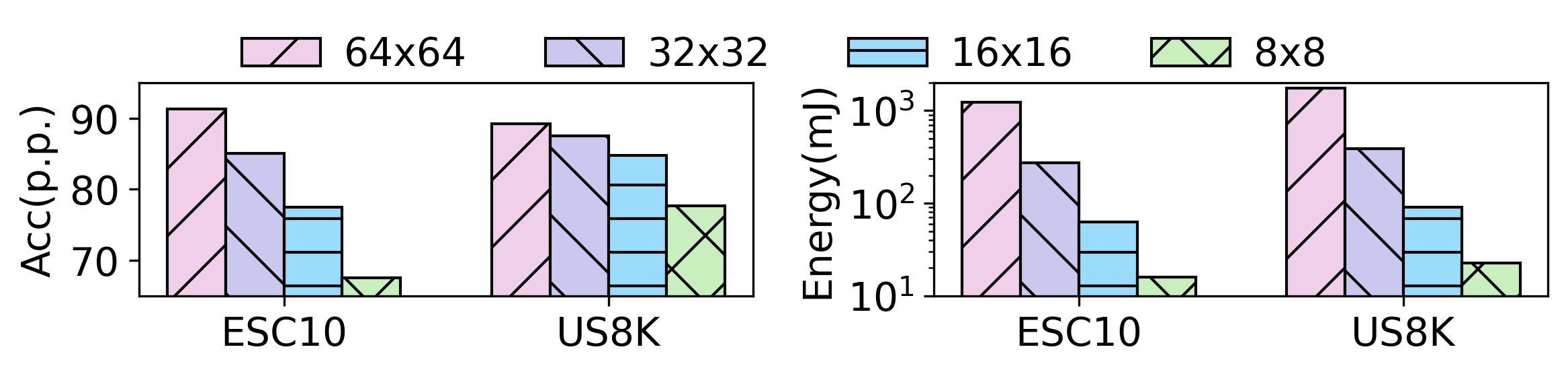}
    \vspace{-0.8cm}
    \caption{\shepherd{Accuracy (left) and energy consumption (right) at various spectrogram resolutions.}}
    \label{fig:resolution-accuracy-energy}
    \vspace{-0.3cm}
\end{figure}

\noindent
\shepherd{ \textbf{Study 1: WPT Depths and Spectrogram Resolution.}
As discussed in Section~\ref{sec:sound-recognition}, the Wavelet Packet Transform (WPT) decomposes signals into finer sub-frequency bands at each level, with spectrogram resolution depending on WPT depth. Greater depth improves classification but increases computational cost. We conduct a preliminary study on WPT depth in environmental sound classification using ESC10~\cite{piczak2015esc} and US8K~\cite{salamon2017us8k}. On an MSP430 microcontroller~\cite{texas2021msp430}, we implemented a simple CNN classifier using WPT spectrograms at varying resolutions, measuring accuracy and energy consumption. Figure~\ref{fig:resolution-accuracy-energy} shows that higher resolution improves accuracy but greatly increases energy consumption, highlighting the need for cost-efficient approaches to balance performance and efficiency. This experiment also implies that to achieve good classification in on-cloud inference, high-resolution spectrogram will need to be transmitted. This results in even larger energy and communication overhead for edge devices, hence motivating keeping the inference pipeline local.}

\begin{figure}[tp]
    \centering
    \includegraphics[width=\linewidth]{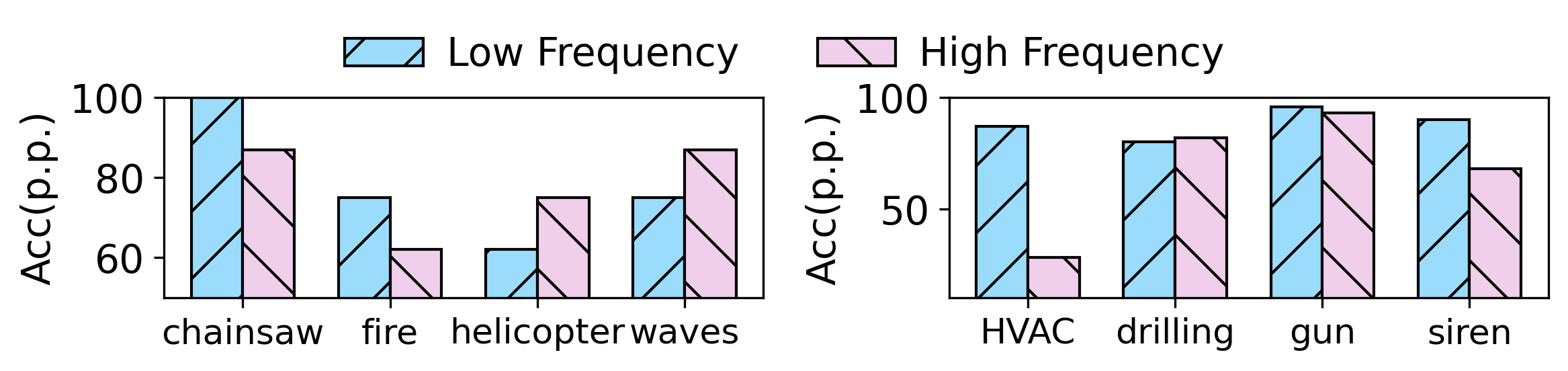}
    \vspace{-0.8cm}
    \caption{\shepherd{Accuracy of using high- and low-frequency band for ESC10 (left) and US8k (right).}}
    \vspace{-0.5cm}
    \label{fig:high-low-frequency}
\end{figure}

\noindent
\shepherd{\textbf{Study 2: Effects of Frequency Bands.} WPT also allows us to selectively upsample frequency-domain resolutions on certain frequency bands. We argue that the discriminative information for different sound classes is distributed differently across different frequency bands.} To verify that, in the second preliminary experiment, we classify spectrograms of the same resolution but with either high-frequency bands only or low-frequency bands only. The results, shown in Figure~\ref{fig:high-low-frequency}, indicate that, for sounds of helicopters, waves, and drilling, high-frequency bands are more important for making the correct classification, whereas low-frequency bands are more important for some other classes.

\shepherd{These observations motivate the use of frequency-domain attention to guide the wavelet transform in generating multi-resolution spectrograms, achieving high accuracy while minimizing WPT and classification costs. This insight informs the design of our novel neural architecture, detailed in the following section.}



\section{ORCA Design}
\label{sec:system-design}

\begin{figure*}[tp]
    \centering
    \includegraphics[width=\textwidth]{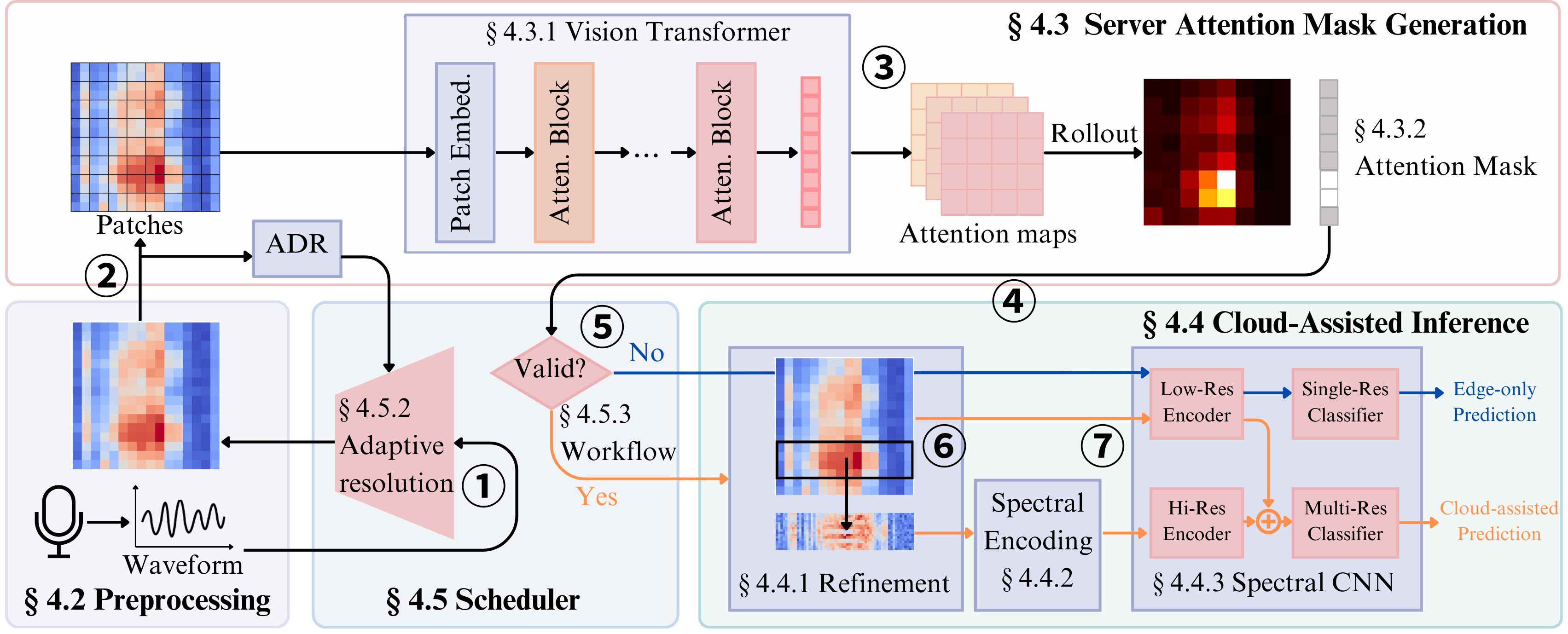}
    \vspace{-0.5cm}
    \caption{\shepherd{ORCA cloud-assisted design overview.}}
    \vspace{-0.2cm}
    \label{fig:system-overview}
\end{figure*}

\subsection{Overview}
\label{sec:system-overview}
Based on the observations and discussions in Section~\ref{sec:background-and-related-works} and~\ref{sec:preliminary-study}, we argue that an ideal edge-cloud collaborative learning system over LPWANs should have the following design considerations. First, to tackle the unreliability of wireless channels, a cloud-assisted strategy should be adopted rather than the state-of-the-art cloud-dependent offloading. Second, to adapt to the low bit rates of LPWANs and the on-device resource constraints, we demand a more efficient information exchange strategy. Additionally, from an audio processing perspective, we look for a more effective feature selection method to reduce input size and therefore reduce on-device computation overheads while maintaining comparable accuracy performances. Informed by these demands, we introduce our novel design of a resource-aware cloud-assisted environmental sounds recognition system, primarily operating over LoRa networks. Our system features resource-aware and communication-adaptive cloud assistance, enabling efficient and flexible cloud offloading under resource constraints and unreliable communications. Furthermore, we apply a novel self-attention-based frequency band feature selection method with the wavelet transform to effectively select important features for efficient on-device inference. We illustrate the workflow of the ORCA cloud-assisted framework in Figure~\ref{fig:system-overview}:

\noindent
\shepherd{\textbf{Step~\textcircled{\small{1}}:} Initially, the edge device preprocesses audio signals using low-level WPT to generate a low-resolution spectrogram. Preprocessing details are in Section~\ref{sec:preprocess}, and optimized resolution selection based on wireless channel feedback, e.g., Adaptive Data Rate (ADR),  is discussed in Section~\ref{sec:resource-aware-cloud-assistance}.}

\noindent
\shepherd{\textbf{Step~\textcircled{\small{2}}:} The resulting low-resolution spectrogram is transmitted to the server via uplink LoRa channel, using ADR-recommended parameters.}

\noindent
\shepherd{\textbf{Step~\textcircled{\small{3}}:} Upon receiving the low-resolution spectrogram, the server processes it using a pre-trained contrastive vision transformer~\cite{dosovitskiy2020vit} to extract an attention mask through attention rollout~\cite{abnar2020quantifying}. Details of the cloud model are provided in Section~\ref{sec:attention-mask-generation}.}

\noindent
\shepherd{\textbf{Step~\textcircled{\small{4}}:} The extracted attention mask, along with ADR feedback, is sent back to the edge device via downlink. Resource efficiency adaptations using ADR feedback are further discussed in Section~\ref{sec:resource-aware-cloud-assistance}.}

\noindent
\shepherd{\textbf{Step~\textcircled{\small{5}}:} The edge device validates the received attention mask. If invalid or lost, it bypasses cloud assistance and performs standalone on-device inference. We will also discuss this in Section~\ref{sec:resource-aware-cloud-assistance}.}

\noindent
\shepherd{\textbf{Step~\textcircled{\small{6}}:} If the mask is valid, the edge device refines the resolution to construct a multi-resolution spectrogram with details in Section~\ref{sec:spectral-encoding-cnn}.}

\noindent
\shepherd{\textbf{Step~\textcircled{\small{7}}:} Finally, with the multi-resolution spectrogram, the edge device performs efficient inference using spectral encoding and spectral CNNs with details in Section~\ref{sec:spectral-encoding-cnn}.}

\shepherd{To address resource constraints and dynamic communication costs, ORCA introduces a novel resource-aware scheduler for efficient cloud assistance on batteryless devices. Our algorithm dynamically adjusts to variable communication costs, enabling optimized communication scheduling in scenarios of high communication costs for adaptive transmission and bypassing. We will detail this algorithm in Section~\ref{sec:resource-aware-cloud-assistance}. }


\subsection{Preprocessing}
\label{sec:preprocess}
\shepherd{To minimize communication costs, ORCA employs a low-resolution wavelet spectrogram as a compact and informative abstraction for cloud assistance. 
We use the WPT with depth $n$ to extract coarse frequency-domain features from the input audio waveform, producing a spectrogram $S$ with a frequency dimension of $2^n$. To generalize features over time and reduce payload size, we apply average pooling along the time axis, transforming $S$ into a square matrix $S_a$ in dimension of $2^n$. We refer to $S_a$ as the cloud-assisted spectrogram and define its dimension as the cloud assistance resolution $R_a = 2^n$, with selection details in Section~\ref{sec:resource-aware-cloud-assistance}.}

\subsection{Attention Mask Generation}
\label{sec:attention-mask-generation}
\shepherd{
In this section, we discuss how the server identifies important features from the assistance spectrogram $\mathcal{S}_a$. Specifically, we define important features as the most informative frequency bands, guided by preliminary studies. The edge device then leverages this information, encoded as an attention mask, to enhance on-device inference accuracy in later steps.
}

\noindent
\subsubsection{Vision Transformer for Assistance Spectrogram.}
\begin{figure}[tp]
    \centering
    \includegraphics[width=\linewidth]{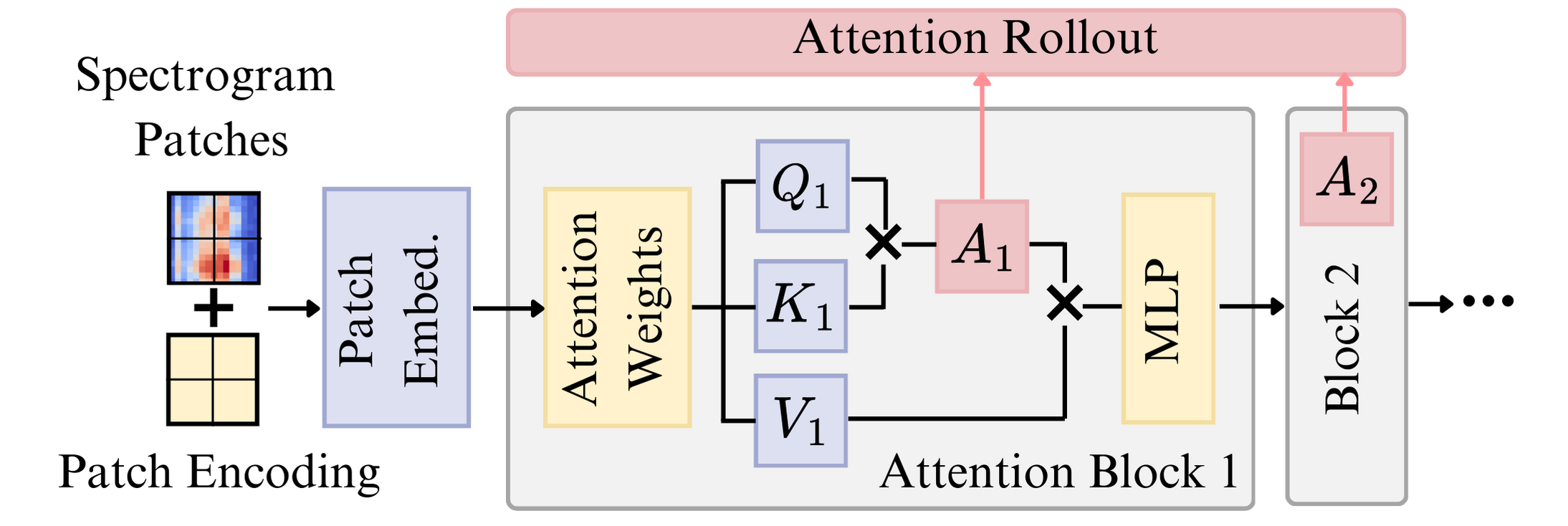}
    \vspace{-0.3cm}
    \caption{Attention computation for attention rollout.}
    \label{fig:attention-block}
    \vspace{-0.3cm}
\end{figure}

\shepherd{ORCA server-side design leverages the self-attention mechanism to dynamically encode the importance of input features. The server processes the assistance spectrogram $\mathcal{S}_a$ by patching it into tokens and computing a self-attention map to highlight key regions. We show the attention computation in Figure~\ref{fig:attention-block}.
First, we adopt the same architecture from the vision transformer~\cite{dosovitskiy2020vit} and divide the input spectrogram into $p^2$ patches. To preserve the spectrogram’s spectral-temporal properties, we apply positional encoding by adding trainable encoding to each patch.
Next, we pass the patches through a convolutional patch embedding layer, encoding each patch into an embedding of dimension $E$.
The resulting embedding is passed through the $i$-th attention block to compute the attention matrix $A_i$, sequentially. Formally, $A_i = \text{Softmax}(Q_i \cdot K_i^T / \sqrt{E})$, where $Q_i$ and $K_i$ are the query and key embeddings at each layer. The attention matrix $A_i$ of size $p^2 \times p^2$ captures the relative importance between patch pairs, aiding in identifying the most informative frequency bands, as discussed next.
}

\noindent
\subsubsection{Attention Mask Generation.}
Recall that the attention matrix $A_i$ represents the attention map of the $i$-th attention block, encoding the relative importance between patches in a spectrogram. 
\shepherd{
Inspired by~\cite{abnar2020quantifying}, we compute the rollout attention map $\widetilde{A} = \Pi^{1}_{i=n} A_i = A_n A_{n-1} \cdots A_{1}$ for importance estimations.} This approach aggregates attention matrices from all blocks, enhancing interpretability and preventing attention scores from vanishing. 
The resulting rollout attention map $\widetilde{A}$ has dimensions $p^2 \times p^2$. 
Then, we aim to identify the most informative frequency bands for the edge. Intuitively, a frequency band is informative if patches within that band have high attention scores, as this indicates that the cloud model prioritizes those patches. Therefore, let $\widetilde{a}_{ij}$ represent the rollout attention between patches $i$ and $j$ in $\widetilde{A}$. We compute the column-wise summation $C$ of $\widetilde{A}$ as $C = [c_1, c_2, \cdots, c_{p^2}]$ where $c_j = \sum_{i=1}^{p^2} \widetilde{a}_{ij}$. 
The vector $C$ is reshaped into a 2D importance matrix $C'\in \mathbb{R}^{p \times p}$, where each entry represents the importance of a patch in the input WPT spectrogram. 
\shepherd{We select frequency bands by summing contiguous $k$ rows in $C'$ and identifying the highest sum, where $k$ is a predefined hyperparameter agreed upon by the server and edge device.} A binary vector of length $p$ records the selected indices, forming the spectral attention mask, which is sent to edge devices.

\noindent
\subsubsection{Contrastive Pre-Training.} \shepherd{The method above relies on a vision transformer capable of identifying informative frequency bands from the WPT spectrogram.} Given the lack of labeled data for frequency-domain feature importance information, we propose training the cloud model offline in an unsupervised manner. \shepherd{Inspired by contrastive learning, where the model learns to produce distinctive features via contrastive loss, we create attracting and contrasting pairs by masking random frequency bands and use triplet loss~\cite{schroff2015facenet} on the flattened output of vision transformer as representations.} Overall, the advantage of ORCA attention-based cloud assistance solution is twofold: first, it uses self-attention over spectrograms to guide clients in focusing on informative frequency bands, which not only improves inference accuracy on the resource-constrained edge devices but also reduces computational load by minimizing the edge model input size. Additionally, transmitting the low-resolution assistance spectrogram and attention masks is highly communication-efficient, significantly reducing communication costs and latency.

\subsection{\shepherd{Cloud-Assisted Inference}}
\label{sec:spectral-encoding-cnn}
\shepherd{
Following the discussion on server-generated attention masks, we explore how edge devices can leverage this information for efficient on-device inference. First, we introduce the \textit{Multi-resolution Refinement} module, which extracts high-resolution frequency bands guided by attention masks. After refinement, two challenges remain: (i) embedding high-resolution spectral bands and (ii) creating a multi-resolution representation for accurate and efficient inference. For (i), we propose \textit{Spectral Encoding}, a trainable weight that encodes high-resolution frequency band-specific knowledge. For (ii), we employ \textit{Multi-resolution CNNs} to process the combination of high-resolution bands from multi-resolution refinement and their corresponding spectral encoding for efficient on-device classification.}

\noindent
\subsubsection{Multi-resolution Refinement.}
\shepherd{The server-generated spectral attention mask captures key frequency bands. It guides the edge device to selectively extract high-resolution spectrograms via wavelet transform. Let $R_l$ denote the pre-defined dimension of the low-resolution spectrogram and $R_h$ the dimension of the high-resolution spectral bands, this refinement results in $R_l$-dimensional low-resolution spectrograms and $R_h$-dimensional high-resolution spectrograms frequency bands.} To further reduce dimension, adaptive average pooling is applied along the time dimension, regularizing the size of both spectrograms.

\noindent
\subsubsection{Spectral Encoding.}
\begin{figure}[tp]
    \centering
    \includegraphics[width=\linewidth]{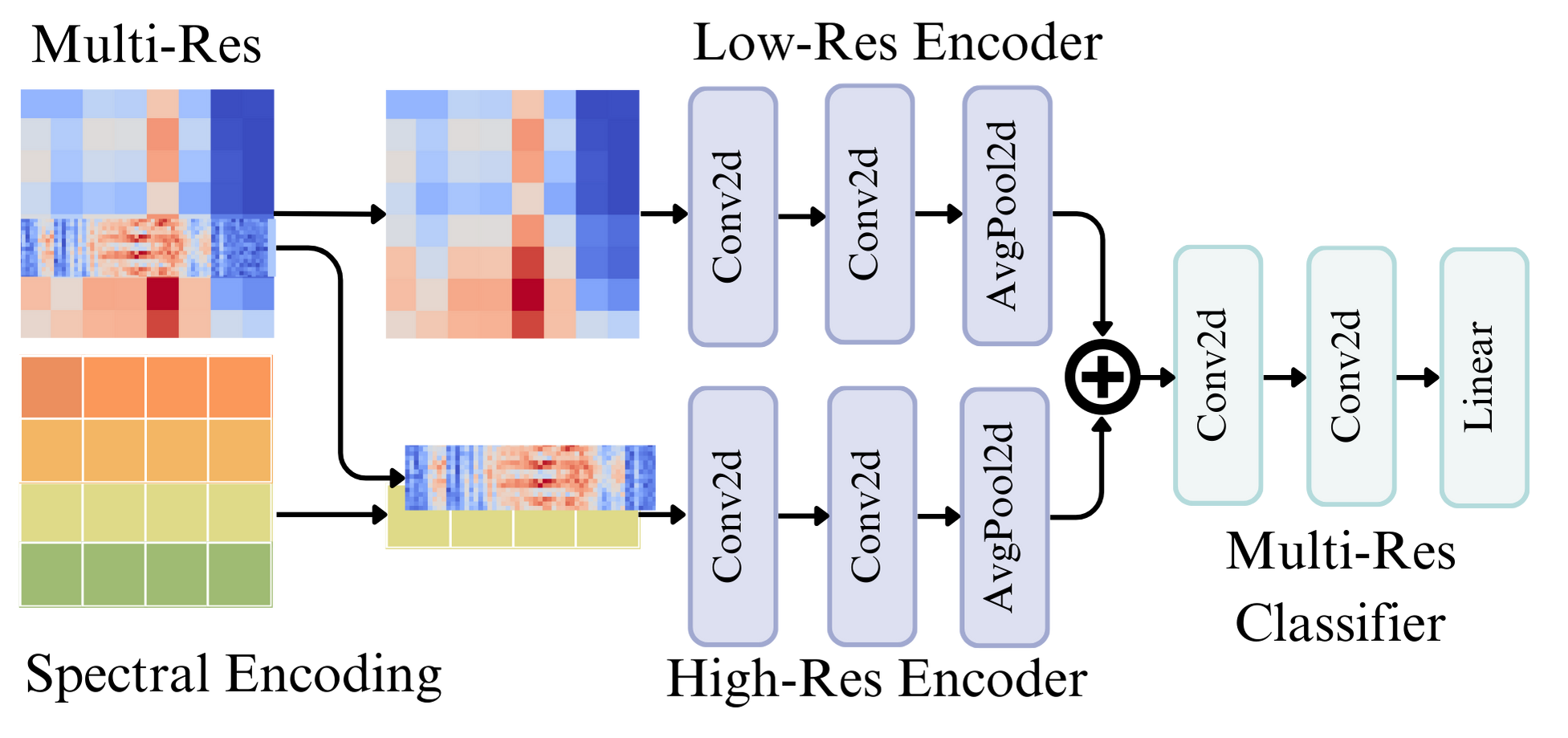}
    \vspace{-0.8cm}
    \caption{\shepherd{Spectral encoding and multi-resolution CNNs.}}
    \label{fig:spectral-encoding}
    \vspace{-0.4cm}
\end{figure}

\shepherd{Since each frequency band captures unique frequency-domain properties, spectrograms from different bands should be interpreted accordingly. Using separate CNNs per band~\cite{phaye2019subspectralnet} is memory-inefficient and costly. Instead, inspired by transformer's positional encoding, we use spectral encoding, a trainable weight that encodes frequency band-specific information. It is then concatenated channel-wise to corresponding high-resolution bands, as shown in Figure~\ref{fig:spectral-encoding}. This approach helps the network to learn spectral-specific knowledge independently of the input spectrogram.}

\noindent
\subsubsection{Multi-resolution CNN.}
\shepherd{The next challenge is to create a multi-resolution representation for inference. As discussed in preliminary studies in Section~\ref{sec:preliminary-study}, discriminative information varies between spectral bands of the spectrogram. With the full low-resolution spectrogram available from preprocessing, we use two 2-layer shallow CNN as encoders, one for low resolution and one for high resolution. The encoded features are fused channel-wise into a single vector and fed into the Multi-Res classifier for final classification. This architecture reduces inference costs by leveraging spectrograms at different resolutions. If cloud assistance is unavailable, an additional Single-Res classifier is employed to process the output of the Low-Res encoder only. All components are pre-trained offline in a two-stage supervised process. First, we train the low-res encoder, high-res encoder, and multi-resolution classifier together with the attention masks generated by the pre-trained cloud vision transformer. In the second stage, we freeze all other components and train the single-resolution classifier independently.}

\begin{figure}[tp]
    \centering
    \includegraphics[width=\linewidth]{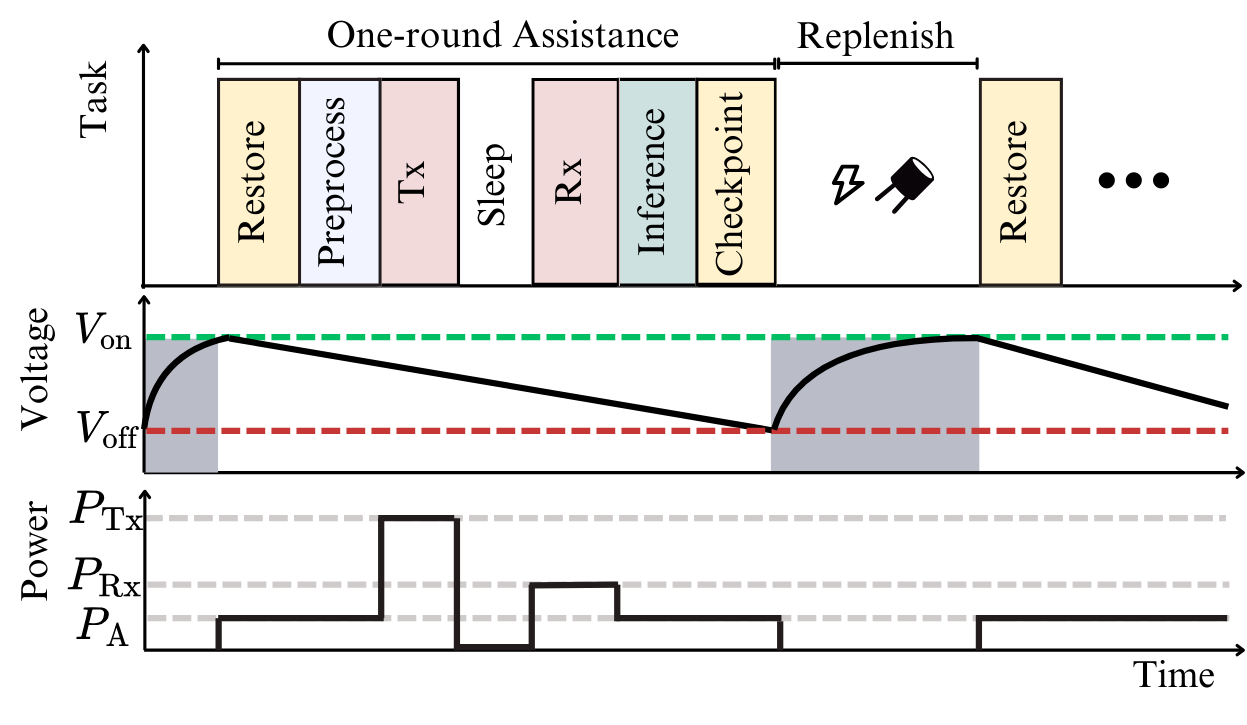}
    \vspace{-1.0cm}
    \caption{Execution model (up), capacitor voltage (mid), and relative power consumptions (low).  }
    \label{fig:intermittent}
\end{figure}

\subsection{Resource-Aware Scheduler}
\label{sec:resource-aware-cloud-assistance}
\shepherd{Given the high energy cost of communication and wireless uncertainty, dynamically managing data transmission size is essential for resource-efficient cloud assistance.} Experimental measurements~\cite{mileiko2023run} indicate that the uplink phase dominates energy consumption in each communication round and varies with channel conditions. Thus, a key component of our framework is optimizing uplink data transmission. 
\shepherd{We introduce a resource-aware, communication-adaptive resolution algorithm. This algorithm dynamically schedules the assistance resolution $R_a$ (as discussed in Section~\ref{sec:preprocess}) based on energy storage and communication quality for resource-efficient cloud assistance.}

\noindent
\subsubsection{Communication Model.} 
As discussed in Section~\ref{sec:system-overview}, ORCA uses two communication phases for one round of cloud assistance, uplink (Tx) and downlink (Rx). It adopts the intermittent computation model from~\cite{mileiko2023run} which concludes an uplink and a downlink in the same power cycle with a synchronized sleep period interleaved. \shepherd{The key advantage of this design is maintaining inference integrity and timeliness for cloud assistance, even during prolonged power failures in batteryless systems.} We illustrate this design in Figure~\ref{fig:intermittent}. Within one power cycle, edge device initiates by restoring the communication parameters, spreading factor (SF) and transmitting power ($P_{\text{Tx}}$) once waking up at voltage threshold $V_{\text{on}}$. Then it goes through sampling and preprocessing, Tx, sleeping, Rx, and on-device inference sequentially as discussed in Section~\ref{sec:system-overview}. Between each power cycle, our edge device checkpoints and restores SF and $P_{\text{Tx}}$ in and out of the non-volatile memory (yellow blocks in Figure~\ref{fig:intermittent}). This ensures their synchronizations to the server's recommendation for reliable communication. Here, the generic ADR algorithm~\cite{Semtech2016LoRaWAN} is employed to estimate the optimal communication parameters ensuring communication reliability. Every time the server receives an uplink packet, it calculates and compares the SNR margins to the optimal values and recommends the optimal SF and $P_{\text{Tx}}$ back to the edge device in downlink message. Edge device can then checkpoint these parameters for next round of communication. The next challenge is to complete restoring, preprocessing, Tx, sleep, Rx, inference, and checkpointing within one power cycle.


\begin{figure}[tp]
    \centering
    \includegraphics[width=\linewidth]{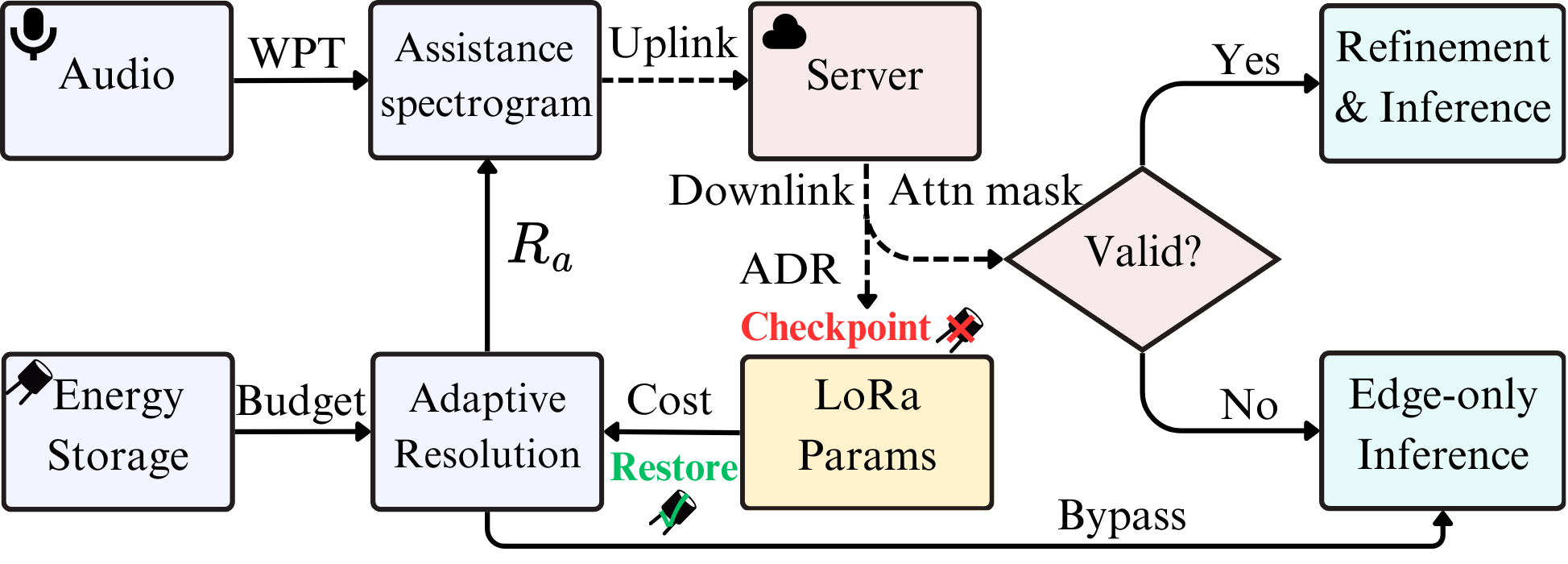}
    \vspace{-0.5cm}
    \caption{Cloud-assisted offloading flow chart.}
    \label{fig:workflow}
    \vspace{-0.2cm}
\end{figure}

\noindent
\subsubsection{Adaptive Resolution.}
\shepherd{Given the proposed communication model and parameters, we first examine the key factors influencing energy consumption.} Since batteryless devices usually wake up at a pre-defined voltage threshold, the energy budget per power cycle is typically fixed and can be estimated by $E_{\text{cap}}=\frac{C}{2}(V_{\text{on}}^2 - V_{\text{off}}^2)$, where $C$ is capacitance, and $V_{\text{on}}$ and $V_{\text{off}}$ represent the microcontroller switching voltage thresholds (on and off, respectively), as depicted in Figure~\ref{fig:intermittent}. 
We propose ORCA resource-aware adaptive resolution algorithm for cloud assistance, designed to adapt to varying communication costs and complete each round of cloud assistance within a single power cycle. Our approach determines an optimal assistance resolution $R_a$ which in turn defines the payload size $S={R_a}^2$ for uplink. We model the adaptive resolution algorithm with the parameters followed. The uplink energy consumption $E_{\text{Tx}}$ can be estimated as: $E_{\text{Tx}} = P_{\text{Tx}} \cdot \text{ToA} = P_{\text{Tx}} ({R_a}^2+S_{\text{p}})/\text{DR}$.
where the uplink transmission time, known as the time-on-air (ToA), depends on the various data rate (DR) under different SF and can be estimated by $\text{ToA} = (S + S_{\text{p}})/\text{DR}$ for sending a payload size of $S$ with a fixed preamble $S_{\text{p}}$. 
The downlink energy cost is estimated as $E_{\text{Rx}}=P_{\text{Rx}} \cdot T_{\text{Rx}}$, the product of the downlink power and the downlink window length. Additionally, $E_{\text{Pre}}$, $E_{\text{sleep}}$, and $E_{\text{inf}}$ are for energy usage during preprocessing, sleep period, and inference, respectively, and can be considered as constants in ORCA. Moreover, to formulate the optimization problem, we define the one-hot encoded resolution selection vector $x$ for resolution ${R_a}$ and the pre-estimated accuracy vector $a$ for accuracy under different ${R_a}$ values. To complete a round of cloud assistance within a single power cycle, the model ensures $E_{\text{Pre}} + E_{\text{Tx}} + E_{\text{sleep}} + E_{\text{Rx}} + E_{\text{inf}} \leq E_{\text{cap}}$. 
We define the following optimization problem, finding the optimal resolution selection vector $x$ to maximize the accuracy under energy constraints:
\begin{equation*}
\begin{aligned}
\max_{x} \ a^{T}x \quad \textrm{s.t.} \quad & E_{\text{Tx}}(x)+E_{\text{Pre}} + E_{\text{sleep}} + E_{\text{Rx}} + E_{\text{inf}}\leq E_{\text{cap}}\\[-0.2em] 
  &\textbf{1}^Tx = 1, \ x_i = \{0, 1\} \\[-0.2em]
  \end{aligned}
\end{equation*}
The optimal resolution selection is derived by ${R_a}=\text{argmax}(x)$, and, specifically, we define ${R_a}=0$ as local bypassing without cloud assistance. In practice, since the optimization search space is small (as $R_a$ is chosen from only a few options) and the capacitor is pre-selected to ensure enough budget for at least local inference without cloud assistance, we simply iterate through all feasible solutions within the energy budget and select the one with the highest estimated accuracy.

\noindent
\subsubsection{Workflow.}
\begin{figure}[tp]
    \centering
    \includegraphics[width=\linewidth]{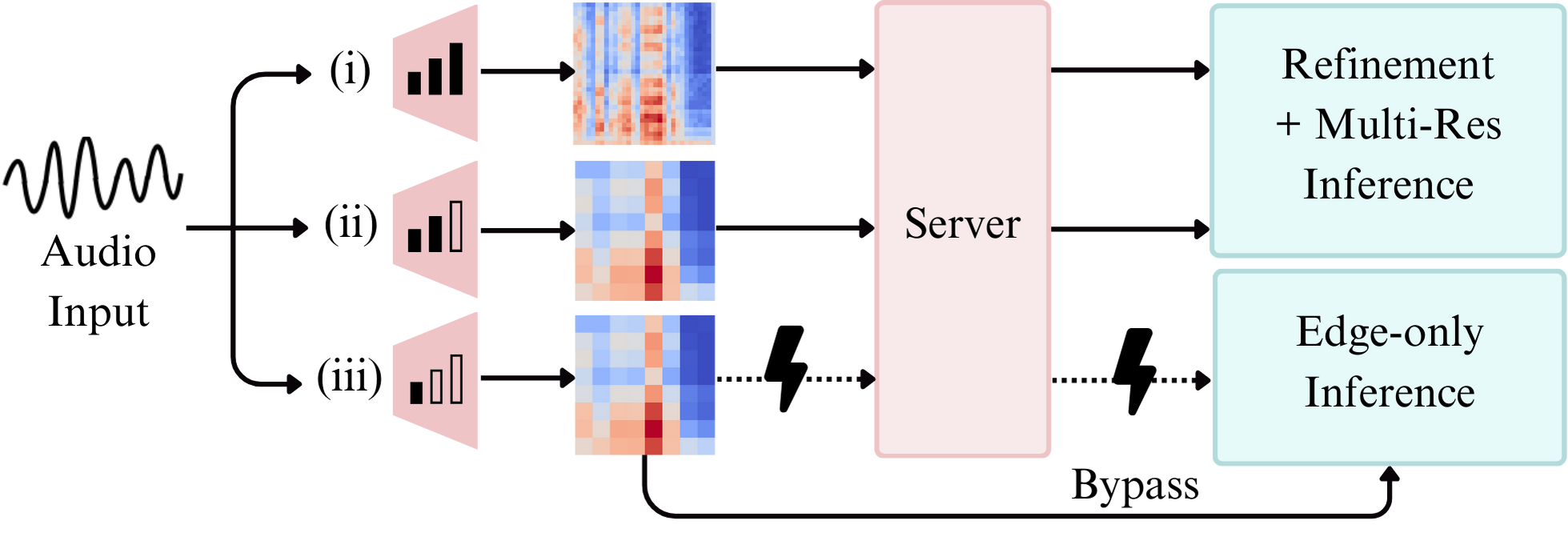}
    \vspace{-0.5cm}
    \caption{Resource-aware adaptive resolution for cloud assistance.}
    \label{fig:resource-aware}
    \vspace{-0.2cm}
\end{figure}

The workflow is presented in Figure~\ref{fig:workflow}. Starting with the communication parameters in the yellow block, we use the energy storage $E_{\text{cap}}$ and communication parameter recommendations from the previous round as the budget and cost inputs, respectively. These inputs are applied to the optimization problem, where the edge device determines the optimal $R_a$ for maximum assistance accuracy and then uploads the low-resolution spectrogram. The server extracts and transmits the attention masks along with the ADR in the downlink back to the edge device. The edge device verifies downlink message validity using the CRC error check or by missing packets after a downlink timeout, treating invalid messages as such. If valid, the edge device proceeds with the multi-resolution inference step as described in Section~\ref{sec:spectral-encoding-cnn}. Otherwise, due to resource constraints, the device bypasses retransmission and cloud assistance, performing single-resolution on-device inference as also detailed in Section~\ref{sec:spectral-encoding-cnn}. Overall, ORCA using fixed energy budgets and dynamic data size offers two major advantages. First, unlike reconfigurable energy storage solutions, which require additional hardware and may face durability or read-write cycle limitations~\cite{colin2018reconfigurable, bakar2022protean, mileiko2023run}, our strategy does not require extra hardware. Second, our algorithm intelligently balances communication costs and accuracy gains by adaptively selecting the amount of resources for cloud assistance. As shown in Figure~\ref{fig:resource-aware}: (i) when communication cost is low, the edge device sends a high-resolution spectrogram for better inference accuracy; (ii) when communication cost is high, it sends a low-resolution spectrogram with a smaller payload to manage energy cost, resulting in lower accuracy; (iii) if communication is unstable with packet loss, the device bypasses cloud assistance and performs local inference to avoid costly retransmissions.

\section{Algorithm Evaluation}
\label{sec:evaluation-on-datasets}

In this section, we assess the effectiveness of ORCA's sub-spectral feature selection algorithm for environmental sound classification using public datasets.

\subsection{Experimental Setup}

\noindent
\textbf{Datasets and Model.}
We evaluate ORCA's model architecture using several public environmental sound datasets, including ESC10~\cite{piczak2015esc}, ESC-nature, ESC-animal, the subsets of ESC50~\cite{piczak2015esc}, US8k~\cite{salamon2017us8k}, and DESED~\cite{turpault2019desed}. The multi-resolution spectral encoding CNN model as discussed in Section~\ref{sec:spectral-encoding-cnn} includes two convolutional layers for each of the low-resolution and high-resolution encoders, and a multi-resolution classifier consisting of two convolutional layers followed by one fully-connected layer.


\noindent
\textbf{Baselines.}
We compare ORCA's attention-based subspectral feature selection strategy with the following state-of-the-art audio feature selection baselines:

\begin{list}{$\bullet$}{\leftmargin=1em \itemindent=0em}
\item \textbf{\textit{Multiscale spectrogram:} } MAST~\cite{zhu2023multiscale} uses fixed resolution spectrogram as audio feature extraction. This includes low-resolution spectrogram on a scale of 16x16 and high-resolution spectrogram on a scale of 64x64. 

\item \textbf{\textit{Time-domain attention selection:} } This implementation uses the attention-based time-domain selection method of SEDAC~\cite{ahn2024split} to identify the most informative audio clips in spectrogram.

\item \textbf{\textit{Frequency-domain amplitude selection:} } This method selects frequency bands with the highest amplitude as in SubSpectralNet~\cite{phaye2019subspectralnet}.
\end{list}

\noindent
\textbf{Metrics.}
For all baseline feature selection methods, classification is performed using a four-layer CNN followed by a fully connected layer. To evaluate ORCA’s effectiveness in extracting the most informative spectrogram features, we compare classification accuracy with state-of-the-art methods under varying $K$ -- the percentage of most informative regions selected. For MAST~\cite{zhu2023multiscale}, we only report accuracy only for high and low resolutions.

\begin{figure}[tp]
    \centering
    \includegraphics[width=\linewidth]{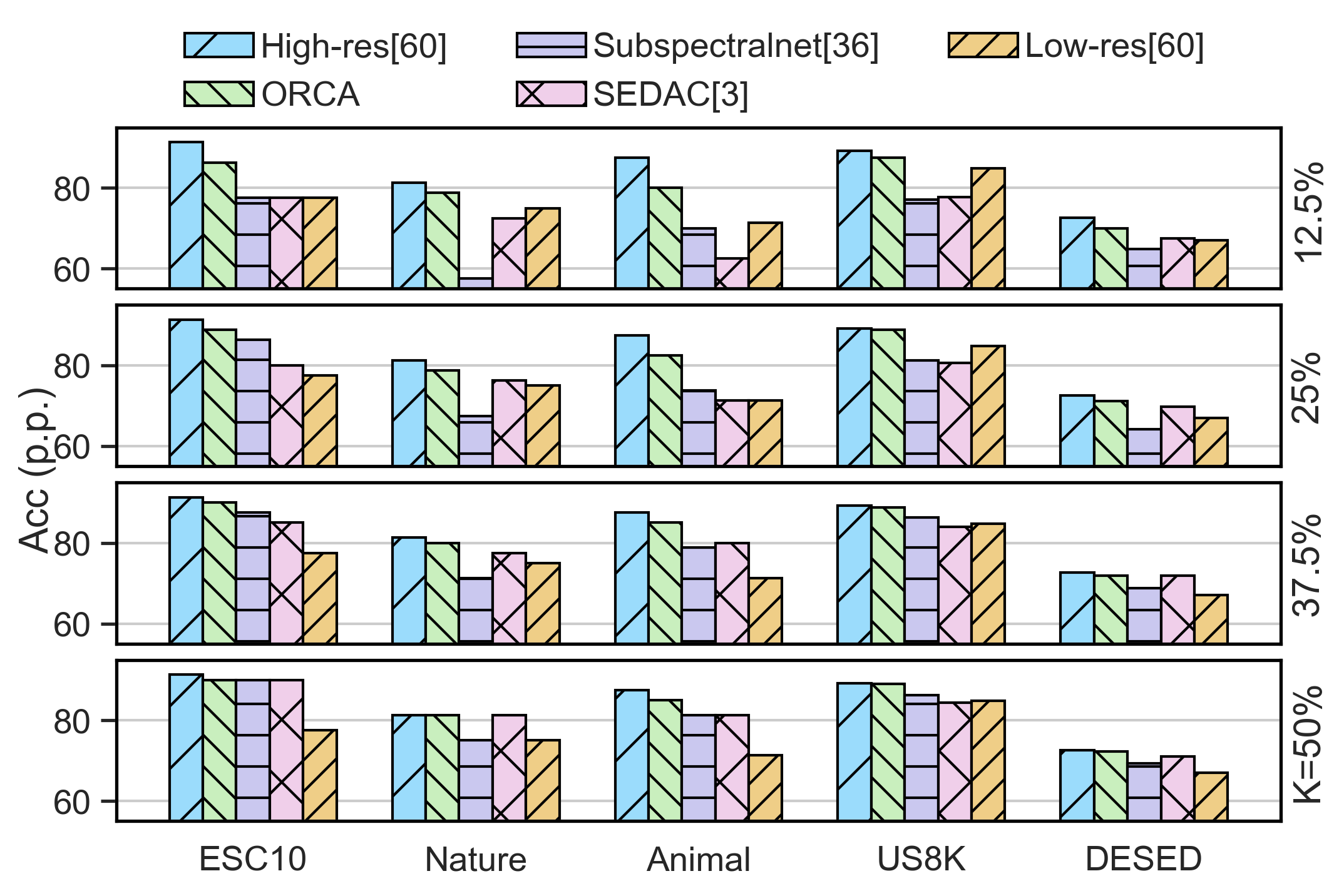}
    \vspace{-0.6cm}
    \caption{Classification accuracy comparisons to the state-of-the-art under various $K$.}
    \label{fig:sota}
    \vspace{-0.4cm}
\end{figure}

\subsection{Comparisons to the State-of-the-art}
\label{sec:comparisons-to-the-state-of-the-art}
Figure~\ref{fig:sota} presents the comparisons for baseline methods and ORCA over five datasets with $K=$ $12.5\%$, $25\%$, $37.5\%$, and $50\%$, as we observe negligible accuracy difference from the high-resolution baseline at $K=50\%$ and higher.
ORCA achieves accuracy closest to the high-resolution baseline, outperforming other efficient baseline methods at each $K$ value. Under extreme resource constraints at $K=12.5\%$, ORCA shows only a 2–5 percentage point (p.p.) accuracy drop compared to the high-resolution baseline, surpassing other baselines by up to 20 p.p. As $K$ increases, ORCA’s accuracy approaches the high-resolution baseline while remaining significantly above other designs. At $K=50\%$, ORCA’s accuracy degradation is just 0.2–2.5 p.p., consistently outperforming other efficient baselines by 2.5–12.5 p.p., demonstrating the effectiveness of ORCA. A key difference between ORCA multi-resolution design and single-resolution methods is in handling less informative regions. Single-resolution methods like SEDAC~\cite{ahn2024split} and SubSpectralNet~\cite{phaye2019subspectralnet} discard these regions entirely. In contrast, our results show that simpler processing of less informative regions, alongside sophisticated processing of the most informative ones, yields superior performance with minimal overhead.

\begin{figure}[tp]
    \centering
    \includegraphics[width=\linewidth]{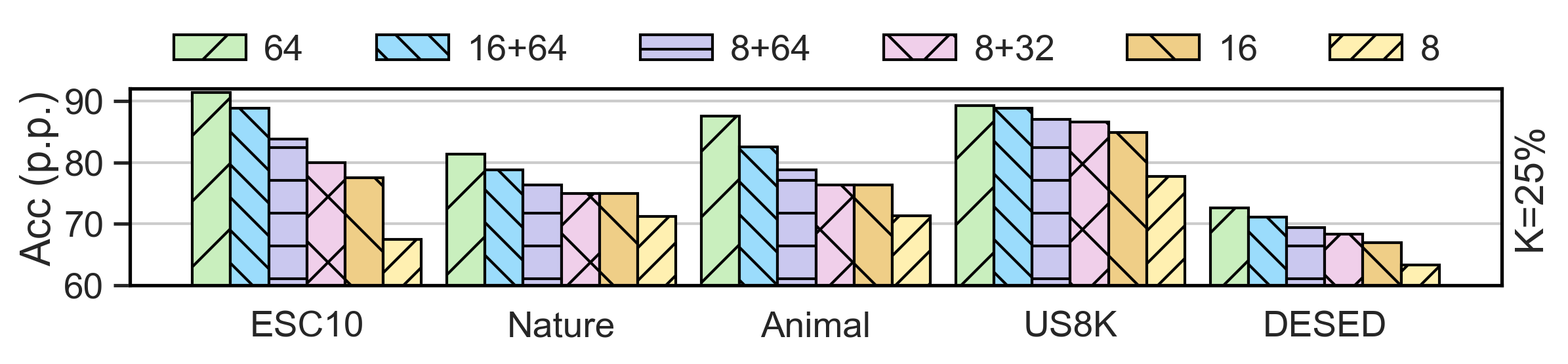}
    \vspace{-0.5cm}
    \caption{Accuracy comparisons of local inference resolutions with top-$K=25\%$ informative regions. }
    \vspace{-0.5cm}
    \label{fig:inference}
\end{figure}

\begin{figure}[tp]
    \centering
    \includegraphics[width=\linewidth]{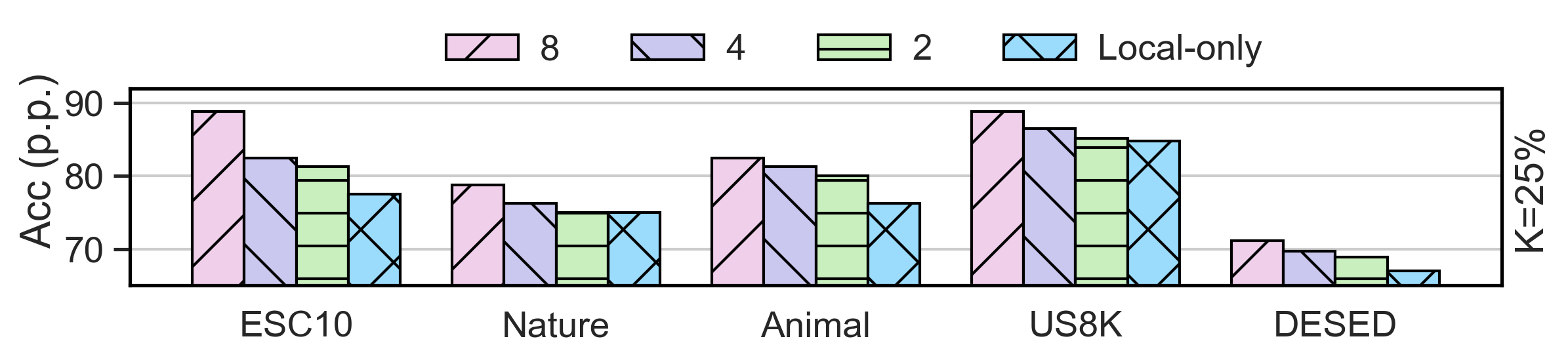}
    \vspace{-0.5cm}
    \caption{Accuracy comparisons of different assistance resolution $R_a$. }
    \label{fig:collaborate}
    \vspace{-0.6cm}
\end{figure}

\subsection{Ablation Study}
\label{sec:sensitivity-analysis}
In this section, to further understand how model parameters influence the classification accuracy, we investigate the key resolution parameters in ORCA with fixed $K$ values: inference resolution ($R_l$ and $R_h$) and assistance resolution ($R_a$) as discussed in Section~\ref{sec:system-design}. In ORCA, the selection of $R_l$ and $R_h$ is pre-defined based on the developer's choices and hardware capabilities. On the other hand, the assistance resolution $R_a$ represents the resolution of the low-resolution spectrogram sent to the server for attention mask generation and determines the cost of communication. This parameter is dynamically adjusted in ORCA system based on communication quality as mentioned in Section~\ref{sec:resource-aware-cloud-assistance}.

\noindent
\textbf{Effects of Inference Resolution ($R_l$ and $R_h$):} 
Figure~\ref{fig:inference} presents the accuracy of various multi-resolution configurations for on-device inference. Single-resolution setups are used for comparison purposes and indicated by the spectrogram dimensions (e.g., 8, 16, 64). Multi-resolution setups are denoted by a combination of $R_l$ and $R_h$, separated by a plus symbol (e.g., 16+64 for $R_l=16$ and $R_h=64$).
The results demonstrate that both $R_l$ and $R_h$ influence the classification accuracy. For example, 16+64 outperforms 8+64 due to the lower low-resolution dimension, while 8+64 outperforms 8+32 due to the lower high-resolution dimension. Overall, our experiments show a relationship between inference resolutions and accuracy. Higher resolutions demonstrate superior performances while requiring more computational resources. On the opposite, lower resolutions show degraded accuracy while being budget-friendly.

\noindent
\textbf{Effects of Assistance Resolution ($R_a$):} 
Figure~\ref{fig:collaborate} presents how the assistance resolution $R_a$ influences the classification accuracy. The number represents $R_a$, and "local-only" is for local on-device inference only without cloud assistance. Our results demonstrate that assistance improves accuracy up to 11 p.p., compared to on-device inference only. Overall, classification accuracy positively correlates to $R_a$ used for cloud assistance, and thus correlates to the packet sizes and energy costs for transmissions. These results will be used as accuracy estimators (i.e. the accuracy vector $a$ in the optimization problem in Section~\ref{sec:resource-aware-cloud-assistance}) storing on edge devices for resource-aware cloud assistance in the subsequent real testbed implementation in Section~\ref{sec:testbed}.

\section{Real-World Testbed Evaluation}
\label{sec:testbed}

\subsection{Testbed Implementation}
\label{sec:testbed-implementation}

\begin{figure}[tp]
    \centering
    \includegraphics[width=\linewidth]{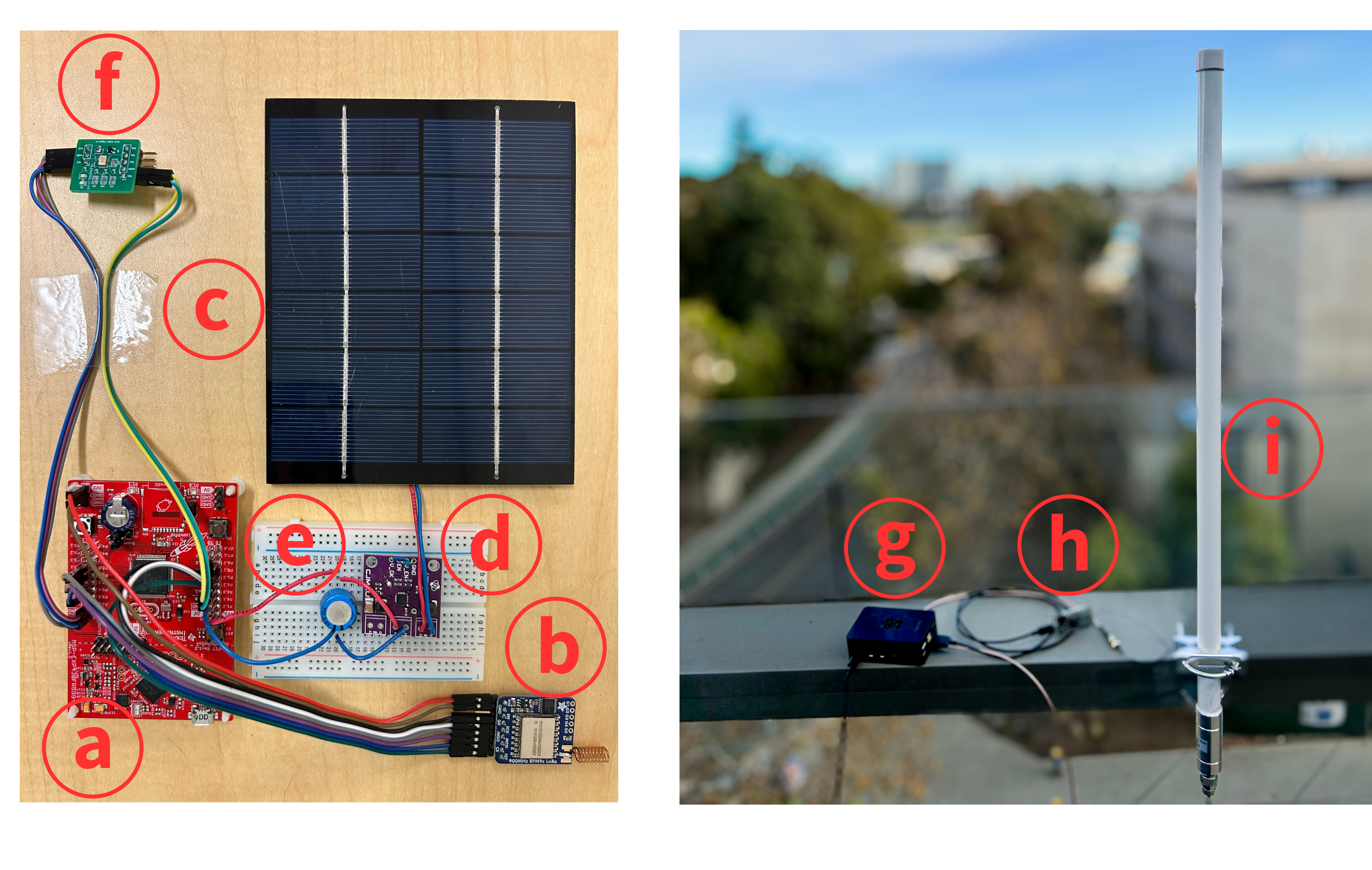}
    \vspace{-0.8cm}
    \caption{Testbed deployment, edge device (left) and server (right). We detail the setup \textcircled{\small{a}}-\textcircled{\small{i}} in Section~\ref{sec:testbed-implementation}.}
    \vspace{-0.2cm}
    \label{fig:real-deployment}
\end{figure}

\noindent
\textbf{Edge device:} We use the 16-bit MSP430FR5994 microcontroller~\cite{texas2021msp430}(\textcircled{\small{a}} in Figure~\ref{fig:real-deployment}) with 8KB of SRAM and 256KB of FRAM, operating at a 16MHz clock frequency, as the edge device. We connect MSP430 to the RFM95W LoRa Radio Transceiver~\cite{hoperf2016RFM95} and a 2dBi spring antenna (\textcircled{\small{b}} in Figure~\ref{fig:real-deployment}). The solar board (\textcircled{\small{c}} in Figure~\ref{fig:real-deployment}) connects to the BQ25570 energy harvester~\cite{ti2019bq25570} (\textcircled{\small{d}} in Figure~\ref{fig:real-deployment}) with a capacitor (\textcircled{\small{e}} in Figure~\ref{fig:real-deployment}), serving as the primary ambient power source for the edge devices. A VM1010 microphone~\cite{vesper2017VM1010} with wake-on-sound technology captures environmental sounds exceeding a predefined amplitude threshold (\textcircled{\small{f}} in Figure~\ref{fig:real-deployment}).

\noindent
\textbf{Server:}
Raspberry Pi 4 Model B~\cite{RaspberryPi4B2019} with 8GB of memory is used as a server to assist the MSP430 edge device (\textcircled{\small{g}} in Figure~\ref{fig:real-deployment}). An Adafruit Feather M0 with RFM95 microcontroller~\cite{adafruit2019FeatherM0} (\textcircled{\small{h}} in Figure~\ref{fig:real-deployment}) with a 5.8dBi fiberglass antenna (\textcircled{\small{i}} in Figure~\ref{fig:real-deployment}) is connected to the Raspberry Pi through the serial port. Both edge and server operate on a 125MHz channel within the spreading factors between 7 to 12 and Tx power from 5 to 17 dBm.

\noindent
\textbf{Software:} The PyWavelets library~\cite{Lee2019PyWavelet} is employed for preprocessing the audio clips, while PyTorch~\cite{paszke2019pytorch} is used for pretraining the server and edge models. Raspberry Pi server also uses PyTorch~\cite{paszke2019pytorch} for cloud assistance. Additionally, we have developed a custom wavelet transform library and an inference engine~\cite{zhang2022demo} integrated with the Low-Energy Accelerator (LEA) on the MSP430~\cite{MSPLEA2016}, aimed at accelerating both preprocessing and inference processes. For the server side, we use the Arduino RadioHead library~\cite{mccauley2013radiohead}. For MSP430, We adapt the RadioHead library~\cite{mccauley2013radiohead, eccob2020msp430} over the SPI interface to facilitate communication with the RFM95W transceiver for both uplink and downlink. To the best of our knowledge, this is the first usable and open-source implementation of the LoRa library for the MSP430. 

\noindent
\textbf{Controlled environment:}
To ensure fair comparisons and reproducibility, we collect real-world data and replicate it in a controlled environment using a two-step testbed. First, we implement an MSP430-based data collector with an RFM95W transceiver and TSL2591 luminosity sensor~\cite{ams2013TSL2591} to gather communication traces (SNR, RSSI), LoRa parameter recommendations (SF, $P_{\text{Tx}}$) , and light conditions from target environments. In the second step, we simulate these environments in the lab using programmable Philips Hue bulbs~\cite{lightbulb2024philips} controlled by the phue Python library~\cite{studioimaginaire2020phue} to replicate lighting, with the server assisting edge inference by simulating communication traces and providing parameter recommendations. Environmental sound clips are played through a speaker.

\subsection{Experimental Setup}

\begin{figure}[tp]
    \centering
    \includegraphics[width=\linewidth]{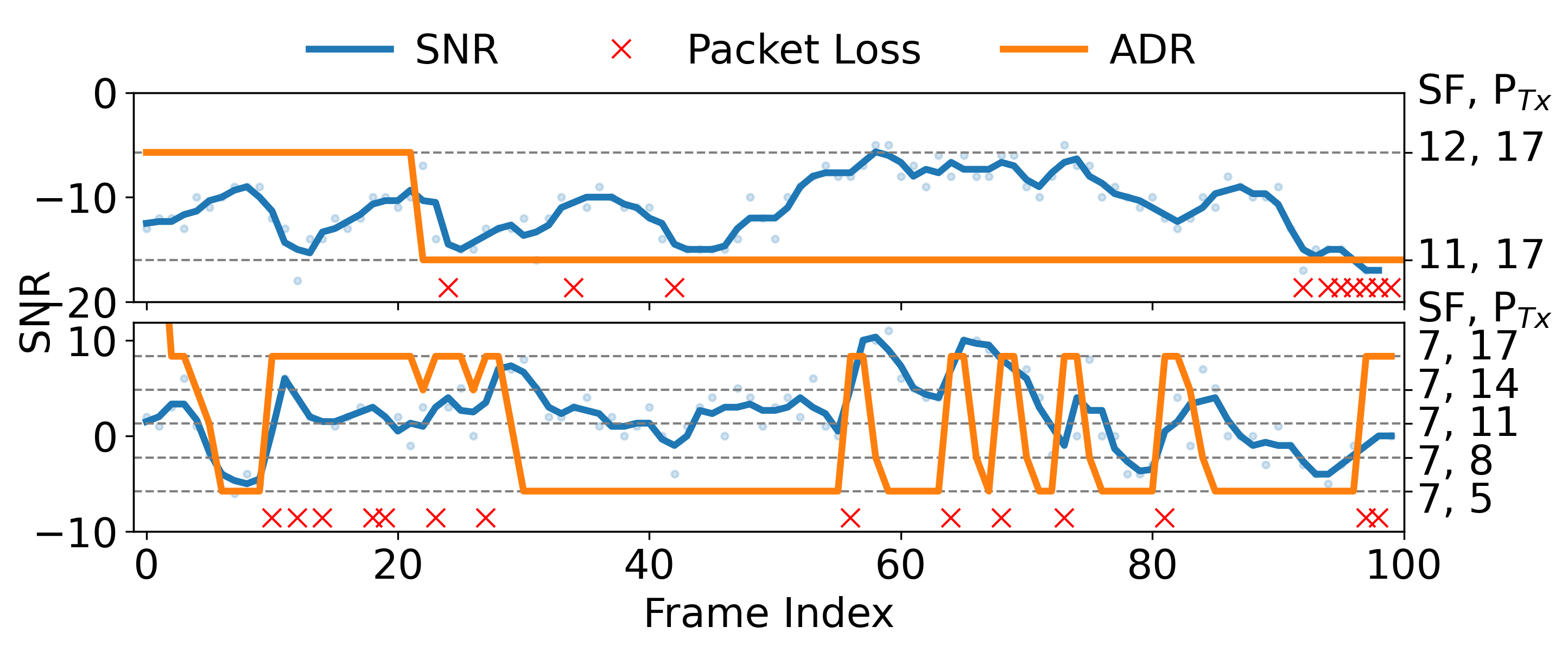}
    \vspace{-0.8cm}
    \caption{Traces of SNR, packet loss, and the ADR for Scenario 1 (up) and Scenario 2 (down). }
    \label{fig:testbed-communication}
    \vspace{-0.3cm}
\end{figure}

\noindent
\textbf{Datasets:} We randomly select environmental sound clips from the US8k~\cite{salamon2017us8k} dataset to simulate real-world audio events. The audios are padded to 4-second clips and sampled at 16kHz. Additionally, we collect real solar energy traces (around 10k lux luminosity in outdoor daylight conditions), and LoRa communication traces, including SNR, communication costs recommended by ADR, and packet losses for two scenarios:
\begin{list}{$\bullet$}{\leftmargin=1em \itemindent=0em}
    \item \textbf{\textit{Scenario 1:}} We set up the server and edge device at a distance of 500m in a complex urban environment with potential obstructions including buildings and moving vehicles. The SNR, packet losses, and ADR traces are presented in top of Figure~\ref{fig:testbed-communication}. Since this scenario requires high energy costs for long-distance wireless communication, we select 100mF capacitor as energy storage.
    \item \textbf{\textit{Scenario 2:}} We set up our system at a distance of 300m in line of sight. The SNR, packet losses, and ADR traces are presented in top of Figure~\ref{fig:testbed-communication}. We select 33mF capacitor as energy storage for low cost short-distance communication. All communication traces are sampled at 1-minute intervals. 
\end{list}

\noindent
\textbf{Baselines:} We summarize the baseline methods as followed:

\begin{list}{$\bullet$}{\leftmargin=1em \itemindent=0em}
\item \textbf{\textit{On-device Inference:}} This implementation adopts the classic on-device inference without cloud offloading~\cite{gobieski2019intelligence,lee2019intermittent} for different resolution spectrograms. 

\item \textbf{\textit{Audio Compression:}} MP3 and AAC~\cite{tomar2006converting} are the two state-of-the-art audio compression algorithms for compressing raw audio waveforms and cloud offloading. This includes different compression bitrates. 

\item \textbf{\textit{Autoencoder Offload:}} DeepCOD~\cite{yao2020deep} and FLEET~\cite{huang2023rethink} compress latent features using an autoencoder before transmitting them to the server. This implementation includes different levels of compression rates. 

\item \textbf{\textit{Progressive Offload:}} Both SEDAC~\cite{ahn2024split} and LimitNet~\cite{hojjat2024limitnet} use context-aware feature selection for offloading. They are combined into one baseline, termed progressive attention-based selection, since LimitNet’s saliency method~\cite{hojjat2024limitnet} is unsuitable for spectrograms. This implementation incrementally includes top-$K$ most informative regions.

\end{list}

\noindent
\textbf{Metrics:} We measure the uplink payload size for cloud assistance, energy consumption, end-to-end inference latency, classification accuracy, and system overhead of ORCA. For energy, latency, and accuracy, we compute the average values across all events in two scenarios respectively.

\subsection{Payload Size}
\label{sec:payload-size}

\begin{figure}[tp]
    \centering
    \includegraphics[width=0.9\linewidth]{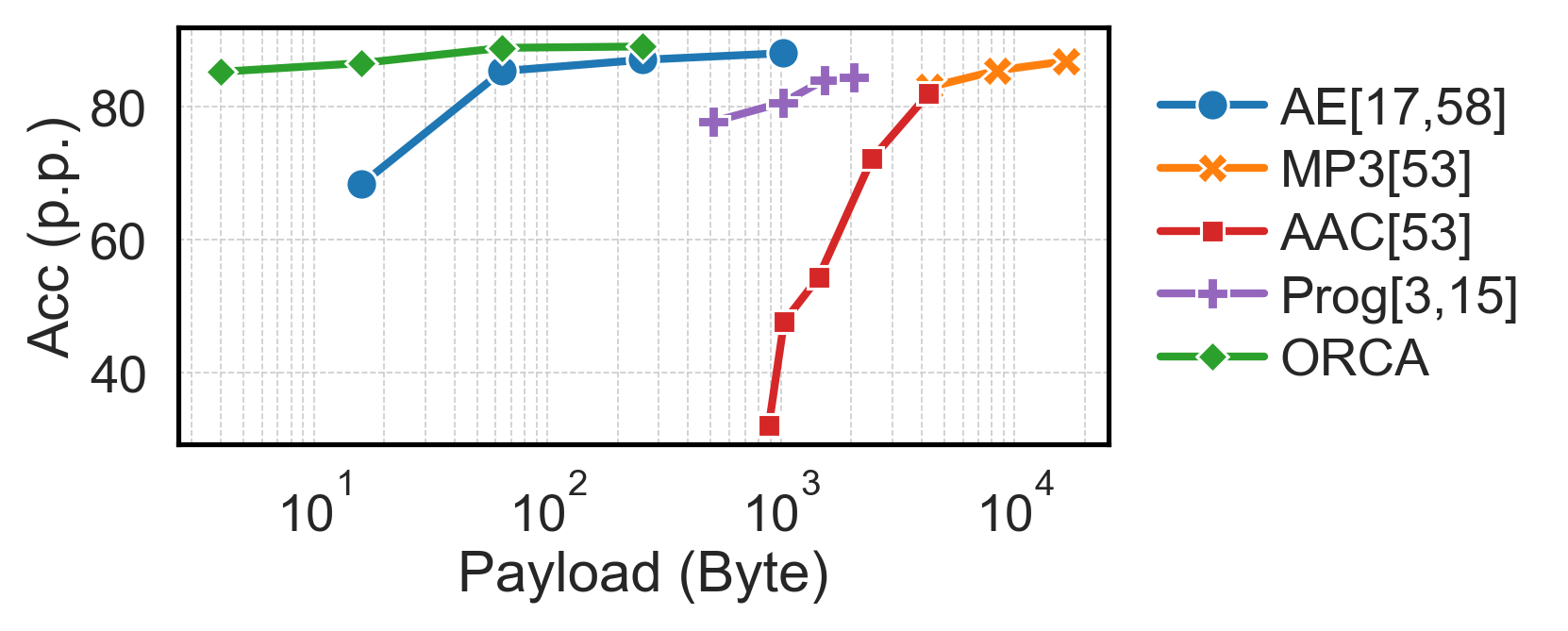}
    \vspace{-0.3cm}
    \caption{Comparisons of accuracy and payload sizes for baseline methods.}
    \label{fig:system-payload}
    \vspace{-0.3cm}
\end{figure}

One of the major benefits of ORCA's cloud assistance strategy we discussed previously is that we can now share much smaller low-resolution spectrogram for feature selection on the cloud, rather than directly use them for high-accuracy cloud inference. Therefore, data compression can be more aggressive than a cloud-dependent strategy without severe accuracy degradation. In Figure~\ref{fig:system-payload}, we compare ORCA to the baselines in terms of accuracy and uplink payload size. First, we notice that the time-series audio compression algorithms MP3 and AAC have poor accuracy even with thousand bytes of payload which is $10\times$-$100\times$ larger than ours. The primary reason is they are generic algorithms for audio compression and are not co-optimized with downstream classification tasks. Next, we notice ORCA outperforms both autoencoder and progressive offloading methods with up to 10 p.p. accuracy advantage and $4\times$-$8\times$ payload savings within the range of 0.1-1 KB, thanks to ORCA's cloud assistance strategy which requires minimal communication for low-resolution spectrograms. Additionally, our method shows a clear 5-20 p.p. accuracy advantage under 100 bytes, the primary operation bitrates under LoRa low data rate mode for long-range transmission~\cite{Semtech2016LoRaWAN}. 

\begin{table}[tp]
    \caption{\shepherd{On-device energy measurement and end-to-end inference latency for each stage. Uplink is a variable for optimization and mentioned in the text.}}
    \centering
    \begin{tabular}{lcccc}
    \hline 
    Metrics & Preprocess & Server & Downlink & Inference  \\
    \hline
    Energy (mJ) & 28.2 & - & 76.5  & 142.0  \\
    Latency (Sec) & 4.0 & 0.02 & 2.0 & 20.1  \\
    \hline
    \end{tabular}
    \label{tab:energy-latency}
    \vspace{-0.3cm}
\end{table}

\begin{figure*}[tp]
    \centering
    \includegraphics[width=\linewidth]{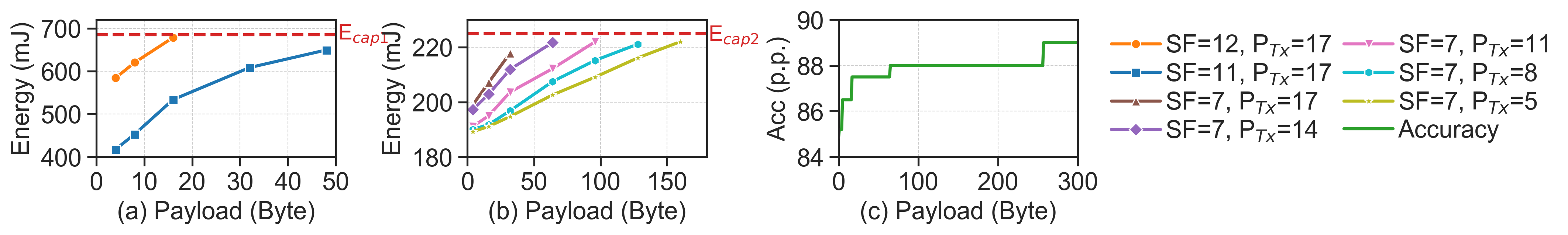}
    \vspace{-0.7cm}
    \caption{The total energy usage for one round of cloud assistance with various uplink payload payloads and LoRa parameters, (a) for Scenario 1 and (b) for Scenario 2. (c) is for accuracy measurement given payloads. }
    \label{fig:system-energy-accuracy}
    \vspace{0cm}
\end{figure*}

\begin{figure*}[tp]
    \centering
    \includegraphics[width=\linewidth]{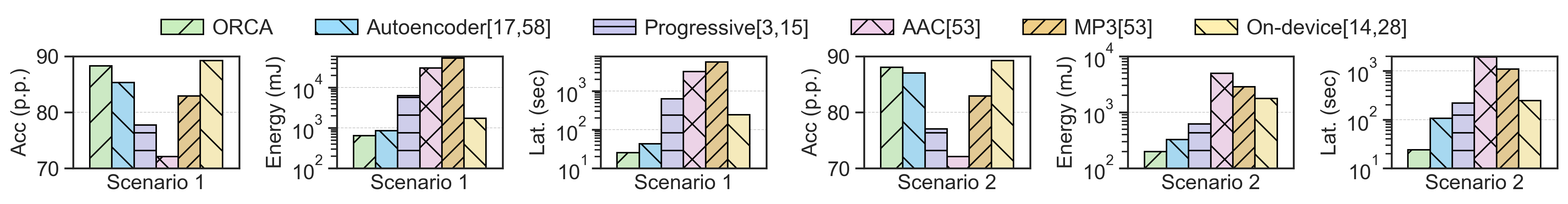}
    \vspace{-0.6cm}
    \caption{The average accuracy, energy, and end-to-end latency for baselines in Scenario 1 (left) and 2 (right).}
    \vspace{-0.2cm}
    \label{fig:scenarios-sota}
\end{figure*}

\subsection{Energy Consumption}

\shepherd{As discussed in Section~\ref{sec:resource-aware-cloud-assistance}, the on-device energy usage in preprocessing, downlink, and inference stages are considered constants in our design. We present measurements of these values in Table~\ref{tab:energy-latency}.} Specifically, preprocessing is done in real-time along with sampling by using an LEA hardware accelerator without time and energy overhead. Additionally, the uplink energy consumption is variable with respect to different payload sizes and LoRa parameters in the optimization problem. Therefore, we measure the total energy consumption of one round cloud assistance as in Section~\ref{sec:resource-aware-cloud-assistance} with respect to the uplink payload size ($S$) under different LoRa parameters in Figure~\ref{fig:system-energy-accuracy}(a) and (b) for Scenario 1 and 2. The results show that with smaller spreading factor (SF) and transmitting power ($P_{\text{Tx}}$) values, ORCA consumes fewer resources as the energy consumption per byte is lower. Given fixed energy budgets $E_{\text{cap1}}=685$mJ and $E_{\text{cap2}}=225$mJ for Scenario 1 and 2 respectively as the red lines in Figure~\ref{fig:system-energy-accuracy}(a) and (b), smaller SF and $P_{\text{Tx}}$ values allow larger payload size $S=R_a^2$ and therefore higher assistance resolution $R_a$. Then, the higher assistance resolution $R_a$ leads to higher feature importance estimation on server side and results in higher cloud assistance accuracy as Figure~\ref{fig:system-energy-accuracy}(c). 
 
Furthermore, we compare the energy consumption of ORCA to the state-of-the-art methods in Figure~\ref{fig:scenarios-sota}. Our results show that ORCA outperforms both autoencoder and progressive offloading by 25\% and 90\% energy savings in Scenario 1 and 40\% and 60\% energy savings in Scenario 2 even with up to 15 p.p. accuracy advantage. The primary reason for these savings is the huge payload savings in ORCA design as we explained in Section~\ref{sec:payload-size}. Additionally, ORCA avoids retransmission which costs additional energy consumptions under resource constraints. ORCA shows a clear advantages of 40$\times$ to 80$\times$ and 8$\times$ to 25$\times$ energy savings in two scenarios compared to MP3 and AAC audio compression algorithms. ORCA saves 64\% and 90\% energy compared to the vanilla on-device inference with merely 1 p.p. accuracy degradation by leveraging the resource-efficient cloud-assisted feature selection.

\subsection{End-to-End Latency}
\shepherd{First, we present measurements of ORCA’s constant latency components, including on-device preprocessing, server processing, downlink, and on-device inference in Table~\ref{tab:energy-latency}. We then incorporate variable uplink latency under two scenarios to evaluate overall end-to-end latency, comparing ORCA with state-of-the-art methods in Figure~\ref{fig:scenarios-sota}.} We notice that given the huge energy usage of LoRa communication, our energy harvesting system takes 50-60 seconds and 10-30 seconds to replenish the energy storage for two scenarios respectively. Compared to autoencoder and progressive baselines, ORCA outperforms both baselines by 40\% and 95\% latency improvements in Scenario 1 and 75\% and 85\% latency improvements in Scenario 2 under accuracy advantages respectively. The primary reason for these improvements is that ORCA avoids the latency of recharging across power cycles by using the resource-aware energy cloud assistance strategy we discussed in Section~\ref{sec:resource-aware-cloud-assistance} to fit all data needed in one round of cloud assistance within one power cycle. Additionally, ORCA avoids retransmission which increases end-to-end with long power cycles. ORCA also achieves 125$\times$ to 220$\times$ and 35$\times$ to 80$\times$ latency improvements in the two scenarios, respectively, compared to MP3 and AAC audio compression algorithms. ORCA saves 64\% and 90\% energy compared to the vanilla on-device inference with merely 1 p.p. accuracy degradation. In both scenarios, ORCA saves 90\% execution time compared to the vanilla on-device inference by the resource-efficient cloud-assisted feature selection.

\subsection{System Overhead}

As discussed in Section~\ref{sec:spectral-encoding-cnn}, we address the overhead of encoding individual frequency bands with a parameter-efficient spectral encoding CNN, using a spectral encoding matrix concatenated with high-resolution inputs. We compared its memory usage to SubSpectralNet~\cite{phaye2019subspectralnet} and a single-resolution on-device model. ORCA’s spectral encoding CNN requires 106KB of non-volatile memory, only an 8\% increase over the single-resolution model’s 98KB, fitting well within the 256KB FRAM of the MSP430. In contrast, SubSpectralNet~\cite{phaye2019subspectralnet} needs 746KB (7$\times$ ORCA’s usage), making it unsuitable for memory-limited microcontrollers.

\section{\shepherd{Discussion and Future Works}}

\noindent
\textbf{Batteryless Computing.}
\shepherd{
ORCA aligns with the batteryless computing paradigm by checkpointing and restoring LoRa communication parameters between power cycles to enable energy-efficient communication. In our design, we treat each round of cloud assistance as an atomic operation for simplicity and timeliness. Beyond that, exploring communication scheduling across multiple power cycles could benefit complex applications. Additionally, we observe that ADR-based parameters may be unreliable under extended duty and power cycles, highlighting the need for a robust algorithm to manage LoRa communication parameters on batteryless devices.}

\noindent
\textbf{Intelligent Acoustic Application.} 
\shepherd{
While ORCA addresses environmental sound classification at the edge, real-world audio-based applications are often more diverse and complex. Exploring areas such as sound event detection, source separation, localization, speech recognition, and integration with human-centric multimodal sensing could further expand its capabilities.}

\noindent
\textbf{Wireless Communication.} 
\shepherd{
Although LoRa and other LPWAN technologies represent state-of-the-art solutions for long-range communication, recent advancements in backscatter offer more energy-efficient designs. Exploring ways to adapt existing LPWAN infrastructure and integrate backscatter radios into batteryless computing devices presents a promising research direction.}


\section{Conclusion}
Environmental monitoring through acoustic signals is increasingly vital across various real-world applications, creating a demand for energy-efficient, environment-adaptive solutions with high model accuracy. In this paper, we introduce ORCA, the first resource-efficient, cloud-assisted environmental sound recognition system on LPWANs, optimized for wide-area audio sensing on batteryless devices. \shepherd{ORCA leverages edge-cloud collaboration by integrating self-attention and vision transformer architectures to extract sub-spectral features in the cloud while exchanging minimal data. This enhances on-device inference accuracy and significantly reduces communication costs necessary for cloud offloading.} As a result, ORCA achieves up to 80$\times$ energy savings and 220$\times$ latency reduction compared to state-of-the-art methods, without sacrificing accuracy.


\begin{acks}
This work was supported in part by the National Science Foundation under Grants \#2112665 (TILOS AI Research Institute), \#2003279, \#1911095, \#1826967, \#2100237, \#2112167, and in part by PRISM and CoCoSys, centers in JUMP 2.0, an SRC program sponsored by DARPA. We thank the reviewers and the shepherd for their insightful comments and suggestions. 
\end{acks}

\balance
\bibliographystyle{acm}

\end{document}